\documentclass[reprint, amsmath,amssymb, aps,prl]{revtex4-2}

\usepackage{graphicx}
\usepackage{dcolumn}
\usepackage{bm}
\usepackage{braket}
\usepackage{color}
\usepackage{stackengine}
\usepackage{ulem}

\newcommand{\rem}[1]{}

\begin{document}

\preprint{APS/123-QED}

\title{Two-Level System Nanomechanics in the Blue-Detuned Regime}

\author{Guillaume Bertel}
\author{Clément Dutreix}
\author{Fabio Pistolesi}

\affiliation{Université de Bordeaux, CNRS, LOMA, UMR 5798, F-33400 Talence, France}

\date{\today}
        
\begin{abstract}
We study a mechanical oscillator coupled to a two-level system driven by a blue-detuned coherent source in the resolved sideband regime. 
For weak mechanical damping, we find dynamical instabilities leading to limit cycles. 
They are signaled by strong fluctuations in the number of emitted photons, with a large Fano factor. 
The phonon-number fluctuations exhibit a strikingly similar behavior. 
When the coupling strength becomes comparable to the mechanical frequency, non-classical mechanical states appear.
We demonstrate that these properties can be detected by measuring the photon-emission spectrum, which enables the reconstruction of the Wigner function.
We then discuss the relation with cavity optomechanical systems. 
Candidates for observing these effects include superconducting qubits, NV centers, and single molecules coupled to oscillators.
\end{abstract}


\maketitle

{\it{Introduction.-}}
Mechanical oscillators coupled to a two-level system (TLS) constitute useful toolboxes for studying fundamental quantum physics and developing quantum technologies.
For instance, driving the TLS at wisely-chosen frequencies permits to manipulate the oscillator, allowing non-thermal state generation \cite{viennot_phonon_2018}, arbitrary quantum states superposition \cite{Hofheinz_Sinthesizing_2009} or ground state cooling \cite{Wilson-rae_laser_2004,Jaehne_laser_2008,rabl_cooling_2010,oconnell_quantum_2010,viennot_phonon_2018}.
Alternatively, the TLS can be used as a probe for the oscillator
displacement \cite{arcizet_single_2012,Bennett_measuring_2012,puller_single_2013,elouard_probing_2019}, 
allowing non destructive measurement of the phonon distribution \cite{ma_nonclassical_2020} and 
force detection \cite{Kolkowitz_coherent_2012,munsch_resonant_2017}.
Reaching a stronger coupling $g_0$ between the oscillator and the TLS leads to larger sensitivities and faster operation. 
When $g_0$ becomes larger than the TLS decay rate $\Gamma$, it becomes possible to manipulate coherently the oscillator with the TLS \cite{Hofheinz_Sinthesizing_2009,Wollack_quantum_2021, Bild_schrodinger_2023}.
Current experimental state of the art is reaching larger values of the 
coupling constant $g_0$ that can be of the same order  
of the mechanical frequency $\omega_m$ \cite{yeo_strain-mediated_2013}.
In these latter systems $\Gamma\gg g_0,\omega_m$, leading to a classical behaviour of the oscillator even at low environment temperature \cite{pistolesi_bistability_2018}. 
However the strong coupling quantum limit $\omega_m \gtrsim  g_0 \gg \Gamma$ remains largely unexplored for the TLS.
In contrast, the quantum limit in cavity opto-mechanics within the same regime has been extensively studied, primarily through numerical methods. 
This research, has led to the prediction of mechanical non-classical states and large phonon fluctuations \cite{ludwig_optomechanical_2008,qian_quantum_2012}. 
%
%
%
Naturally, one might ask whether similar effects could be observed for TLS and how these effects compare to those observed in cavities.
%
%

%
In this paper we consider an oscillator whose displacement induces a modulation of the energy splitting of a TLS \cite{Jaehne_laser_2008,puller_single_2013,pikkalainen_cavity_2015,munsch_resonant_2017}.
We focus on the system's dynamics when the TLS is driven by a coherent field near the first sideband in the blue-detuned regime. 
By choosing a suitable basis, we derive a Pauli rate equation.
This greatly simplifies the understanding of the problem and its solution: the coherences vanish and all the information on the stationary state is contained in the populations of the mechanical Fock states in the energy eigenstate basis. 
We find that by increasing the drive intensity, the system undergoes a transition from a thermal state to a series of limit cycles characterized by a non-monotonic population distribution. 
This transition is marked by large fluctuations in the photons emitted by the TLS, indicated by giant Fano (or Mandel) factors.
This is reminiscent of predictions for the Franck-Condon blockade in quantum transport, for a two-state quantum dot \cite{Grifoni_Driven_1998,koch_franck-condon_2005,Wabnig_Statistics_2005}.
Varying $g_0$ we predict the emergence of mechanical non-classical states with negative Wigner function.
Remarkably, these non-classical quantum states are completely described by the population distribution. 
This stems from the fact that these states resemble Fock states, which, apart from the vacuum state, are inherently non-classical.
We propose a method to extract the Wigner function from the measurement of the photon-emission spectrum.
Finally, we discuss how these results also extend to cavity optomechanics under analogous conditions.

{\it{Theoretical Model.-}}
\begin{figure}
\centering
\includegraphics[scale=0.24]{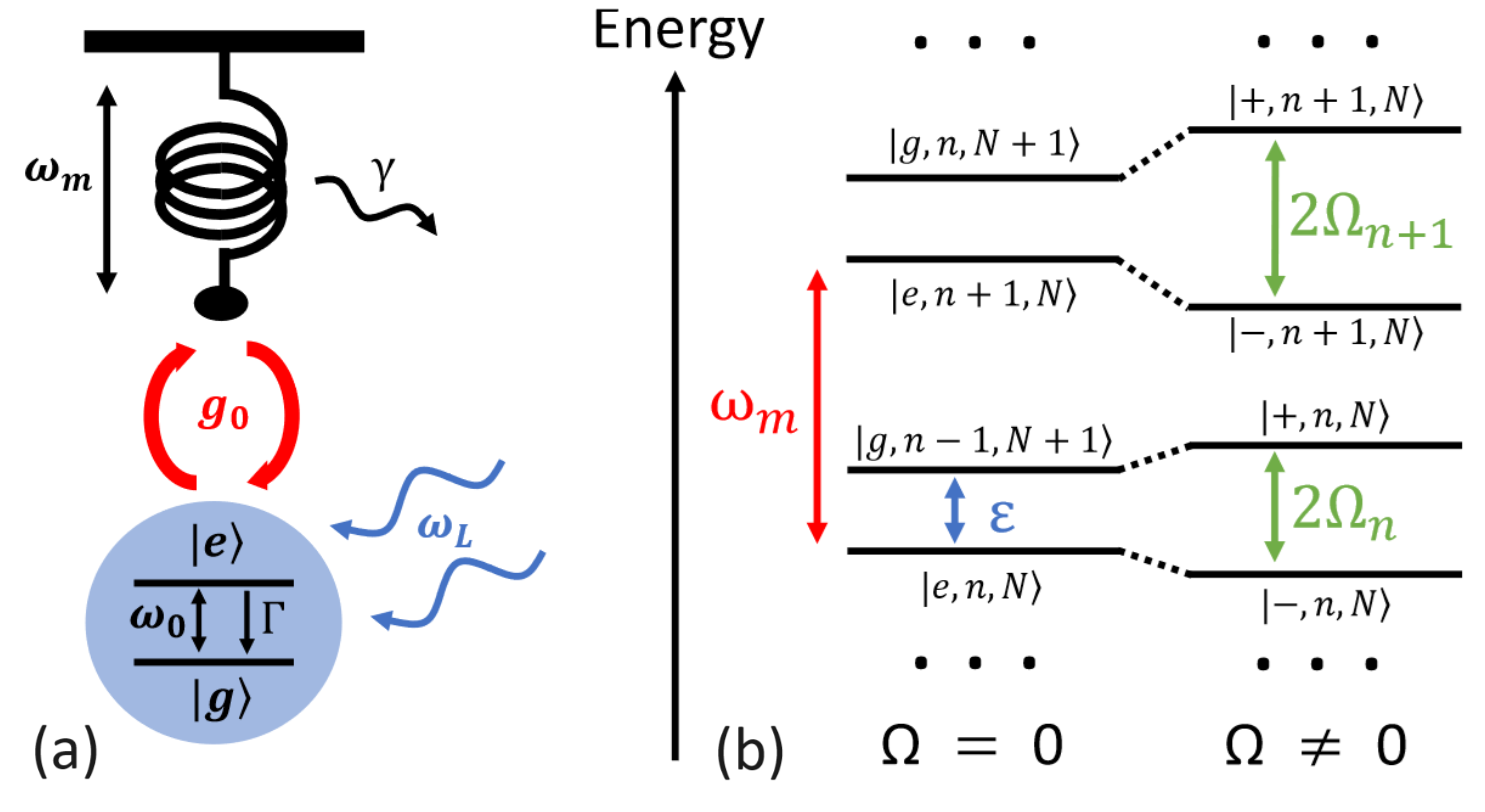}
\caption{
(a) : Schema of the TLS of energy splitting $\omega_0$ driven by a laser of frequency $\omega_L$ and coupled via $g_0$ to an oscillator of resonant frequency $\omega_m$. %
Here $\ket{e}$ and $\ket{g}$ indicate the two states of the TLS, while $\Gamma$ and $\gamma$ are the damping rates of the TLS and the oscillator. (b) : Energy diagram of the system in the dressed atom picture, with only one multiplicity $N$ displayed. Here $\Omega$ is the Rabi frequency, $\ket{n}$ and $\ket{N}$ are the number states of the oscillator and the laser source. The $\ket{\pm,n,N}$ are the eigenstates of the full diagonalized Hamiltonian, splitted by $2\Omega_n$.}
\label{Schema_and_diag}
\end{figure}
We consider a TLS of energy splitting $\omega_0$ (we use $\hbar = 1$) coupled to a single mode of a mechanical oscillator of frequency $\omega_m$ [Fig.~\ref{Schema_and_diag}(a)].
The Hamiltonian of the isolated system reads
\begin{equation}
\label{H_s}
H_s = \omega_m a^\dagger a + \frac{\omega_0}{2} \sigma_z + g_0 (a + a^\dagger) \sigma_z\ . 
\end{equation}
Here $a$ is the annihilation operator for the oscillator mode and $\sigma_i$ is the $i$-Pauli matrix. 
The energy of the TLS is modulated by the mechanical motion, for instance due to Stark or Zeeman effect.
This is relevant for various experimental systems, such as singles molecules \cite{puller_single_2013}, nitrogen vacancy in diamond nano-crystals \cite{arcizet_single_2012} or superconducting qubits \cite{pikkalainen_cavity_2015}.
Note that a vast body of literature exists on TLS in atoms coupled to cavities \cite{Jones_Photon_1999,Clemens_Nonclassical_2000,Hamsen_Two_2017,Kebler_Emergent_2019,Gao_Self_2023}. However, these systems are typically coupled via a Rabi (transverse) interaction term $\sigma_x$ and typically the cavity frequency is close to the TLS energy splitting.
The TLS and the mechanical oscillator are coupled to the electromagnetic (EM) and the mechanical environments through the coupling $\sigma_x E_1$ and $(a+a^\dag) E_2$, respectively, where $E_1$ and $E_2$ are operators of the environments (namely electromagnetic fields and phonon displacements).
For the TLS, this leads to the decay rate $\Gamma = \tilde{C}_1(\omega_0)$, where $\tilde{C}_1(\omega)$ is the Fourier transform of $C_1(t) = \braket{E_1(t)E_1(0)}$ the self-correlation function of the EM environment.
Similarly the oscillator damping is described by $\gamma (n_B+1) = \tilde{C}_2(\omega_m)$, with $C_2(t) = \braket{E_2(t)E_2(0)}$.
Here $n_B=(e^{\omega_m/k_B T}-1)^{-1}$ is the Bose distribution at the mechanical environment temperature $T$ and $k_B$ is the Boltzmann constant.
For the numerical simulations in the following we assume $k_B T = \hbar \omega_m$, which corresponds, for example, to a temperature of $15\ \mathrm{mK}$ (achievable in a dilution refrigerator) for a mechanical frequency $\omega_m = 32 \times 2\pi \ \mathrm{MHz}$.
This implies $n_B = 0.58$.
We assume that the Bose occupation at the EM frequency $\omega_0$ is negligible.
The Hamiltonian $H_s$ can be diagonalized exactly by applying the Lang-Firsov \cite{Lang_transformation_1962} (or polaron) transformation: $H_1 = U_1^\dag H_s U_1 = \omega_m a^\dag a + (\omega_0/2)\sigma_z$, where $U_1 = e^{-g_0(a^\dagger -a)\sigma_z/\omega_m}$.
In the Born-Markov approximation, one can derive a master equation for the system density matrix $\rho_s$ in this diagonal basis.
In the resolved sidebands limit ($\Gamma, \gamma \ll \omega_m$) we find $d\rho_s/dt = \mathcal{L}\rho_s$ (see SM), with
\begin{align}
\label{Dissip_diag}
\mathcal{L} & \rho_s  = -i [H_1, \rho_s] + \gamma (n_B + 1) \mathcal{D}(a)\rho_s + \gamma n_B \mathcal{D}(a^\dagger) \rho_s  \notag \\
& + \Gamma \mathcal{D} \left( \sigma_- e^{-\frac{g_0}{\omega_m}(a^\dagger-a)} \right) \rho_s + \gamma_\phi \mathcal{D}(\sigma_z) \rho_s \ ,
\end{align}
and $\mathcal{D}(O)\rho_s = O \rho_s O^\dagger - (O^\dagger O \rho_s + \rho_s O^\dagger O)/2 $. 
Note that the last term represents a pure dephasing rate for the TLS: $\gamma_\phi = (2g_0/\omega_m)^2 \tilde{C}_2(0)$, which arises from mechanical dissipation mediated by the coupling $g_0$. 
The rate $\gamma_\phi$ is controlled by the zero frequency correlator, it is thus in principle different from the 
damping rate.
For simplicity, we assume $\gamma_\phi = \gamma$ in the following analysis.
Any intrinsic dephasing rate of the TLS can be accounted for by the value of $\gamma_\phi$.
In the Supplementary Material, we show that even when $\gamma_\phi$ is on the order of $\Gamma$, the results presented below remain valid.

\textit{Weak-drive dressed states.-}
We assume that the TLS is driven by a coherent source of frequency $\omega_L$ and intensity $\Omega$.
In the spirit of the dressed atom picture \cite{Cohen-Tannoudji_interaction_2001}, we describe this by a cavity of resonant frequency $\omega_L$ populated by a large number of photons.
The drive Hamiltonian reads
\begin{align}
\label{H_L}
H_L = \omega_L b^\dagger b + \frac{\Omega}{2} \left[ \sigma_+ e^{2\frac{g_0}{\omega_m}(a^\dagger-a)} b + \rm{h.c}. \right] \ ,
\end{align}
where $b$ is the annihilation operator for the photons in the cavity.
In the resolved sideband regime, photons can be absorbed only if $\omega_L -\omega_0 \approx n \omega_m$, where $n$ is an integer.
In the following we consider that the laser is tuned at one of the blue detuned sidebands and focus in particular on the first one by assuming $\omega_L = \omega_0 +    \omega_m + \epsilon$, with $|\epsilon| \ll \omega_m$.

We describe the interaction with the cavity field using a perturbation theory in $\Omega$.
For $\Omega = 0$, the eigenstates of $H_1 + H_L$ are $\ket{\sigma,n,N}$, where $\sigma = \{ e, g \}$ indicates the TLS ground or excited state and $n$ ($N$) labels the phonon (photon) number state.
The eigenvalues read $E_{\sigma,n,N} = \lambda_\sigma \omega_0/2 + n \omega_m + N \omega_L$, with $\lambda_e = -1$ and $\lambda_g=1$.
For the chosen value of the cavity frequency, the states $\ket{e,n,N}$ and $\ket{g,n-1,N+1}$ form nearly degenerate doublets with energy splitting $\epsilon$ [see Fig.~\ref{Schema_and_diag}(b)].
For $\Omega\neq0$, degenerate perturbation theory leads to the dressed eigenstates 
\begin{align}
    \ket{\pm,n,N} = \alpha_\pm (n) \ket{e,n,N} + \beta_\pm(n) \ket{g,n-1,N+1} \ ,
\end{align}
where we introduced $\alpha_{\pm}(n) = \mp\beta_{\mp}(n) = \sqrt{ (\Omega_n \mp \epsilon)/(2\Omega_n)}$ and the Rabi splitting $\Omega_n = \sqrt{(\Omega W_{n,n-1})^2 + \epsilon^2}$ [see Fig.~\ref{Schema_and_diag}(b)].
Here $W_{n,m} = \braket{n|e^{2g_0(a^\dagger-a)/\omega_m}|m}$ is the Franck-Condon factor.
The condition $\omega_m \gg \Omega_n$ ensures that no other state mixes significantly with the doublets.

\textit{Pauli rate equation.-}
The dressed states basis can be used to solve the master equation (\ref{Dissip_diag}), after the inclusion of the drive.
For $\epsilon \gg \Gamma, \gamma$, the secular approximation applies since the energy splitting of the doublets $\Omega_n$ is much larger than $\Gamma, \gamma$.
This simplifies greatly the solution of the master equation, that reduces to the Pauli (rate) equation for the populations of the perturbed eigenstates, while the off-diagonal elements vanish in the stationary limit.
The transition rates given by Eq.~(\ref{Dissip_diag}) for the EM environment from state $\ket{\mu,n,N}$ to state $\ket{\mu',n',N-1}$ read 
\begin{align}\label{TLSRate}
    \Gamma_{\mu,n,N \to \mu',n',N-1} = \Gamma \left| \alpha_\mu(n)\beta_{\mu'}(n') W_{n,n'-1} \right|^2 \ ,
\end{align}
where $\mu = \{+,-\}$.
Similarly, we find the rates induced by the mechanical environment
\begin{align}
\label{Rate_mec}
    &\gamma_{\mu,n,N \to \mu',n\pm1,N} = \left| \alpha_{\mu'}(n\pm1)\alpha_\mu(n)\sqrt{n+\xi_\pm} \right. \notag \\
    & \left. + \beta_{\mu'}(n \pm 1)\beta_\mu(n)\sqrt{n-\xi_\mp} \right|^2 (\xi_\mp + n_B)\gamma \ ,
\end{align}
where we defined $\xi_\pm=(1\pm1)/2$.
In the following, we investigate the stationary state obtained from the Pauli rate equation.

\textit{Limit cycles.-} 
\begin{figure}
\includegraphics[scale=0.215, trim = 2.2cm 0cm 2cm 2cm, clip]{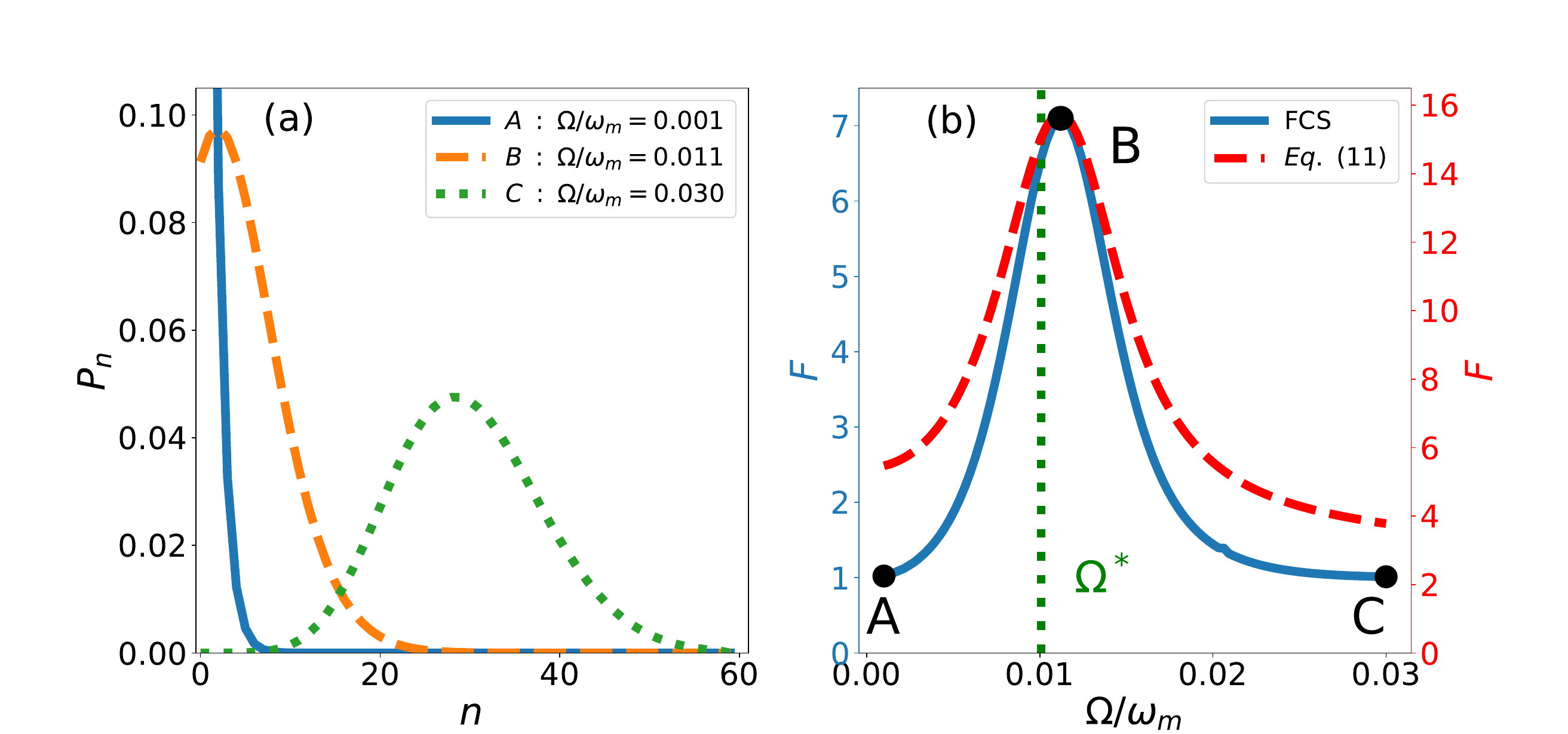}
\caption{
(a) Phonon probability distribution $P_n$ for three different driving intensities labeled as $A$, $B$, and $C$.
(b) Fano factor of the light emitted by the TLS (blue solid line, left vertical axis) compared with the 
approximate Eq.~(\ref{Bruit}) derived from the phonon-number correlation function (red dashed line, right vertical axis),
as a function of the drive intensity $\Omega/\omega_m$. 
The green dotted line refers to the critical drive intensity $\Omega^*$ obtained from Eq. (\ref{n_bar}) and marking the mechanical transition between the thermal steady state and the limit cycle.
In both pictures, the parameters in units of $\omega_m$ are: $g_0 = 0.1$, $\Gamma = 0.01$, $\gamma = \gamma_\phi = 10^{-4} $, $\epsilon = 0.01$.}
\label{Fano_and_Pn}
\end{figure}
Solving numerically the Pauli rate equation we obtain the steady-state phonon population $P_n$ as a function of the drive intensity $\Omega$.
This is shown in Fig.~\ref{Fano_and_Pn}(a) for the coupling strength $g_0/\omega_m = 0.1$.
For weak driving ($\Omega/\omega_m=10^{-3}$), $P_n$ follows a thermal Boltzmann distribution.
We observe that increasing $\Omega$ leads to a peaked distribution with a maximum at a finite value of $n$.
This indicates the appearance of a limit cycle in phase space, similar to the instabilities discussed for cavity optomechanics \cite{Braginsky_Parametric_2001,Rokhsari_radiation_2005,qian_quantum_2012}.
The mechanical transition toward a limit cycle can be understood from the evaluation of the mean number of phonons $\bar{n} = \sum_n n P_n$.
From a general conservation argument, the mean-phonon number obeys the equation of motion.
\begin{align}
\label{n_dot}
    \dot{\bar{n}} = -\gamma(\bar{n}-n_{B})+\Gamma_{\rm op} \ .
\end{align} 
The first term on the right-hand side represents the standard thermalization of an oscillator, and is obtained when expanding at the lowest order in $g_0/\omega_m$ Eq.~(\ref{Rate_mec}) in the rate equation.
The last term describes variation of the number of phonons induced by the emission of photons in the optical environment.
It can be written as 
$\Gamma_{\rm op} = \sum_{p=-\infty}^\infty p I^{(p)}$,
where $I^{(p)}$ is the photon flux 
associated with the emission (or absorption) of $p$ phonons by the mechanical oscillator.
This allows us to introduce the total flux of photons as $\bar{I} \equiv \sum_{p=-\infty}^\infty I^{(p)}$.
It is clear that $\Gamma_{op}$ and $\bar{I}$ are two different quantities.
However, in the limit $g_0\ll\omega_{\rm m}$ we can show that $I^{(1+k)} \simeq I^{(1-k)}$ (see SM).
This implies that the optical mean-phonon rate is equivalent to the total photon flux:
$\Gamma_{\rm op} \simeq \bar{I}$.
In such a way, \textit{on average}, each photon emitted by the TLS is associated with the transfer of one phonon from the TLS to the mechanical oscillator.
This stems from the fact that the TLS is driven in the first blue sideband.
In the steady state ($\dot{\bar{n}}=0$), this leads to the remarkable identity
\begin{equation} 
\label{Flux_1}
\bar{I} \simeq \gamma (\bar{n}-n_B) \ .   
\end{equation}
It links the photon flux to the average number of phonons through the mechanical damping coefficient. 
We verified numerically the validity of Eq. (\ref{Flux_1}) [cf. Fig.~S2(a) in SM].
We can further determine in the limit $g_0 \ll \omega_m$ the photon flux explicitly from the rates of the Pauli Eq.\,(\ref{TLSRate}).
It gives 
\begin{equation}
\label{Flux_2}
 \bar{I} = \Gamma A (\bar{n} + 1) \ ,  
\end{equation}
where $A = \left[\Omega g_0 / (\epsilon \omega_m) \right]^2$ (see SM). 
From Eqs.\,(\ref{Flux_1}) and (\ref{Flux_2}) we finally obtain the mean phonon number of the steady state 
\begin{equation}
\label{n_bar}
 \bar{n} = \frac{\Gamma A + \gamma n_{\rm B}}{\gamma -\Gamma A}  \ .
\end{equation}
It diverges for the critical drive intensity $\Omega^* \simeq |\epsilon | \omega_m \sqrt{\gamma}/(g_0\sqrt{\Gamma})$, signaling the transition toward the limit cycle for the oscillator.
We investigate now the photon fluctuations.
We can readily obtain the full counting statistics of the emitted photons from the Pauli description of the dynamics.
For this, we introduce a counting field $\chi$ in the rate equation \cite{Bagrets_full_2003,pistolesi_full_2004,bennett_full_2008,xu_full_2013} (see SM). 
The solution of the equation, $P_{i,n}(\chi,t)$, gives the generating function of the emitted photons ${\cal S}(\chi,t) = \ln [ \sum_{i,n} P_{i,n}(\chi,t)]$, where $i$ labels the state of the system.
From $\cal S$ we obtain the photon flux $\bar{I} = \partial \dot{\cal S}_{t\to\infty}(\chi,t) / \partial (i \chi) |_{\chi \to 0}$ and the zero frequency noise $S_{\rm II} = \partial^2 \dot{\cal S}_{t\to\infty}(\chi,t) / \partial (i \chi)^2 |_{\chi \to 0}$, where $\dot{\cal S}_{t\to\infty}$ is the time-derivative of ${\cal S}$ evaluated in the limit of $t \to \infty$.
These two quantities are readily measured in experiments by using photocounters.
They provide a measure of the photon fluctuations through the Fano factor $F = S_{II}/ \bar{I}$ or, equivalently, the Mandel factor $Q=F-1$.
A Fano factor of $1$ is an indication of Poissonian statistics.
We represent in Fig.~\ref{Fano_and_Pn}(b) (blue solid line) Fano factor $F$ of the photon as a function of the drive intensity $\Omega$.
It is larger than 1, which indicates super-Poissonian statistics with the emission of photons in bunches.
We find that it can take very large values ($F\gg1$) at the transition point $\Omega^*$.
This shows that the mechanical instability toward limit cycles is accompanied with strong fluctuations in the photon statistics.

Equation (\ref{Flux_1}) suggests that the photon flux fluctuations are related to the phonon number fluctuations in a simple way $\delta I \sim \gamma \delta n$.
From this observation one can expect then
\begin{equation}
\label{Bruit}
    \frac{S_{\rm II}}{\bar{I}} \approx \gamma \frac{S_{\rm nn} - S_{\rm nn}^{\rm th} }{\bar{n}-n_B} \ ,
\end{equation}
where $S_{nn}$ indicates the zero-frequency noise spectrum of the phonon number, and where we subtracted its thermal value so that for $\Omega=0$, the photon fluctuations correctly vanish.
We resort to the quantum regression theorem and evaluate numerically the right-hand side of Eq. (\ref{Bruit}), which we compare to the photon Fano factor in Fig.~\ref{Fano_and_Pn}(b) (red dashed line).
This shows that within a numerical factor  $\approx 2.3$, Eq.~(\ref{Bruit}) holds.
The large Fano factor of the photon flux can thus be attributed 
to the strong phonon-number fluctuations.
This marks the transition toward the self-oscillation regime.

{\it{Non-Classical States.-}} 
\begin{figure}
\centering
\includegraphics[width=1 \columnwidth]{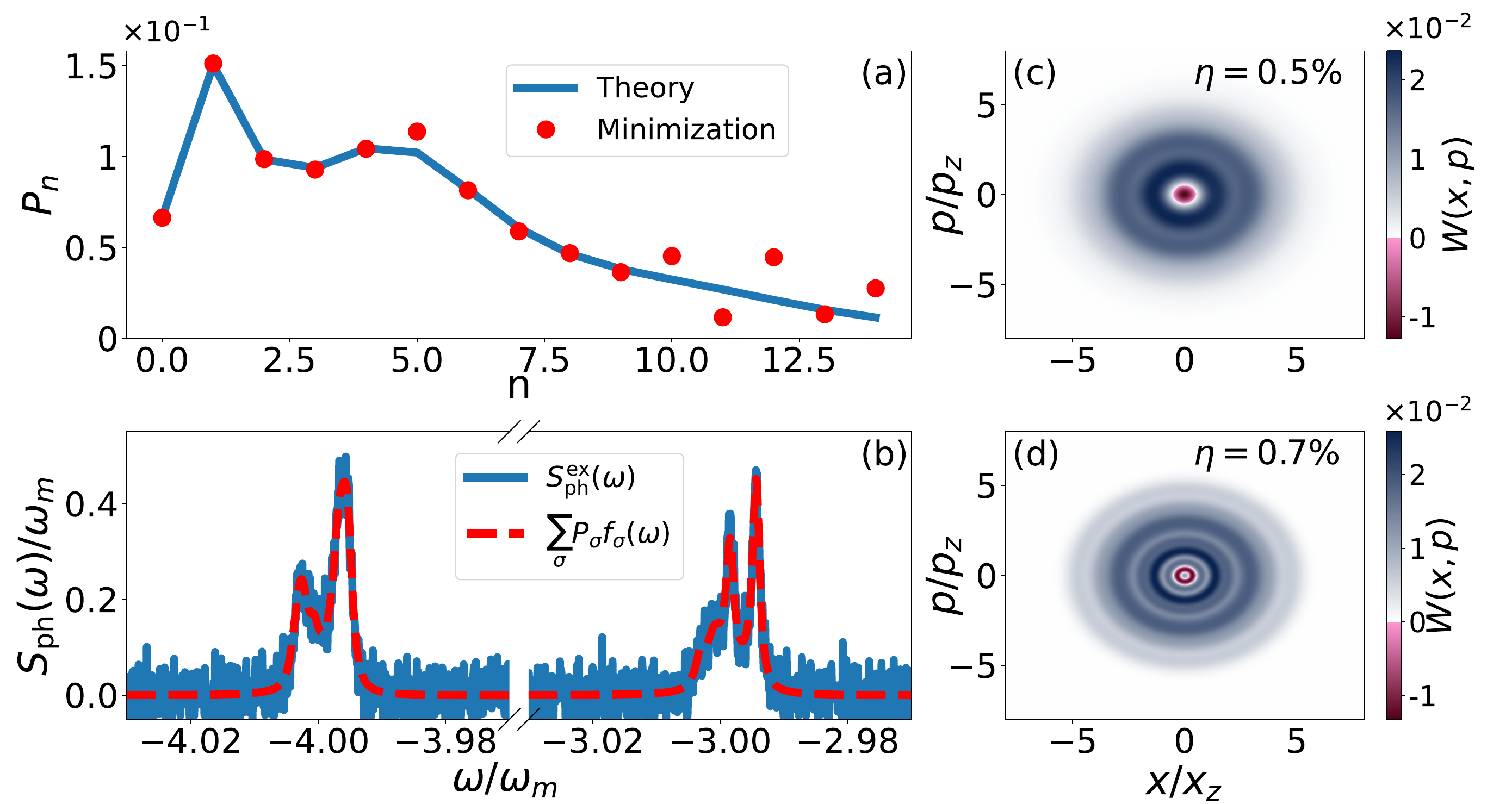}
\caption{(a) Phonon probability distribution $P_n$ computed from the master Eq.~(\ref{Dissip_diag}) (blue solid line) and from the fit to the noisy spectrum of inset (b).
(b) Photon-emission spectrum around two side-bands, obtained by the resolution of the master equation and the addition of a random noise (blue solid line), compared to the result of the fit with Eq.~(\ref{Spectrum}) (red dashed line). 
(c) Wigner function obtained from the resolution of the master equation.
(d) Wigner function obtained from the fit to the spectrum.
The parameters, in unit of $\omega_m$, are : $g_0 = 0.7$, $\Omega = 0.25$, $\Gamma = 0.01$, $\gamma = \gamma_\phi = 10^{-4}$, and $\epsilon = 0.3$. }
\label{Tomography}
\end{figure}
Increasing the drive intensity $\Omega$ or the electro-mechanical coupling $g_0$, additional maxima in the phonon distribution $P_n$ appear [cf.~Fig.~\ref{Tomography}(a)].
From $P_n$ it is relatively straightforward to obtain the Wigner function $W(x,p) = \sum_n P_n W_n(x,p)$, where $W_n(x,p) = [(-1)^n/\pi] e^{-(x^2+p^2)} \mathcal{L}_n[2(x^2+p^2)]$ is the Wigner distribution associated to the number state $\ket{n}$, and $\mathcal{L}_n$ is the Laguerre polynomial of order $n$ \cite{Groenewold_1946}.
Remarkably, we find that the Wigner function can be negative [see Fig.~\ref{Tomography}(c,d)], indicating the formation of a non-classical state.
The non-classicality can be traced back to the sharpness of the distribution $P_n$, which occurs for instance for the first peak in Fig.~\ref{Tomography}(a).
This state is very similar to a single Fock state, $P_n = \delta_{n,n_0}$, with $n_0 \neq 0$, which has a negative Wigner function \cite{lorch_laser_2014}.
To characterize the negativity of $W(x,p)$, we introduce the factor $\eta = -\int_- W(x,p) dx dp / \int_+ W(x,p) dx dp$, where $\int_\pm$ refers to the integral over the phase-space where the Wigner distribution is negative ('-') or positive ('+') \cite{Nation_Nonclassical_2013}.
In Fig.~\ref{Eta} we show $\eta$ for various values of $\Omega$ and $g_0$.
We see the appearance of fringes for nearly constant values of $g_0$, for which the state is non-classical. 
Each fringe correspond to a sharp peak at different values of $n$, as indicated in the figure, with $n$ increasing by reducing the coupling $g_0$. 
The typical maximum value of negativity, as displayed on Fig.~\ref{Eta}, is $\eta \approx 1 \%$.
\textit{Measurement of the negativity} -
We propose a method to experimentally measure the density matrix of the mechanical part, which, in the secular approximation, reduces to the populations.
This information is sufficient to extract the full Wigner function and obtain the negativity of the non-classical state.
For the parameter range we have considered here, this can be achieved by the measurement of the photon-emission spectrum $S_{\rm ph}(\omega) = 2 {\rm Re} \int_0^{+\infty} dt e^{i \omega t} {\rm Tr}\left[ \sigma_+^d e^{\mathcal{L}t}\sigma_-^d \rho_{\rm st} \right]$.
Here $\sigma_\pm^d$ is the Pauli operator in the diagonal basis, and $\rho_{\rm st} = \sum_\sigma P_\sigma \ket{\sigma}$ represents the steady-state density matrix, with $\ket{\sigma} = \ket{\pm,n,N}$.
This implies
\begin{equation}
\label{Spectrum}
S_{\rm ph}(\omega) = \sum_{\sigma} P_\sigma f_\sigma(\omega) \ ,
\end{equation}
where $f_\sigma(\omega) = 2 {\rm Re} \int_{-\infty}^{+\infty} dt e^{i \omega t} {\rm Tr} \left[\sigma_+^d e^{\mathcal{L}t} \sigma_-^d \ket{\sigma}\bra{\sigma} \right]$.
The strong nonlinearity of the problem results in a frequency dependence of $f_\sigma(\omega)$ that varies significantly with $\sigma$.
Using linear minimization methods to fit experimental spectra, we expect that it should be possible to extract $P_\sigma$, and, in turn, the phonon probability distribution $P_n$ (see SM).
To show the feasibility of this method, we simulate an experimental spectrum [$S_{\rm ph}^{\rm ex}(\omega)$] by calculating the emission spectrum [$S_{\rm ph}(\omega)$] and adding to each point [$S_{\rm ph}(\omega_i)$] a Gaussian random noise with mean $S_{\rm ph}(\omega_i)$ and standard deviation $0.03\ \omega_m^{-1}$.
Fig.~\ref{Tomography}(b) shows a comparison between $S_{\rm ph}^{\rm ex}(\omega)$ and Eq.~(\ref{Spectrum}) where the $P_n$ have been obtained by fitting $S_{\rm ph}^{\rm ex}(\omega)$.
The minimization allows to retrieve remarkably well the predicted distribution $P_n$ [see Fig.~\ref{Tomography}(a)], and consequently the Wigner distribution [Fig.~\ref{Tomography}(c,d)].
We discuss the method's robustness against noise and sampling rate variations in the SM.
\begin{figure}
\centering
\includegraphics[scale=0.22, trim = 3cm 0cm 6.75cm 1.75cm, clip]{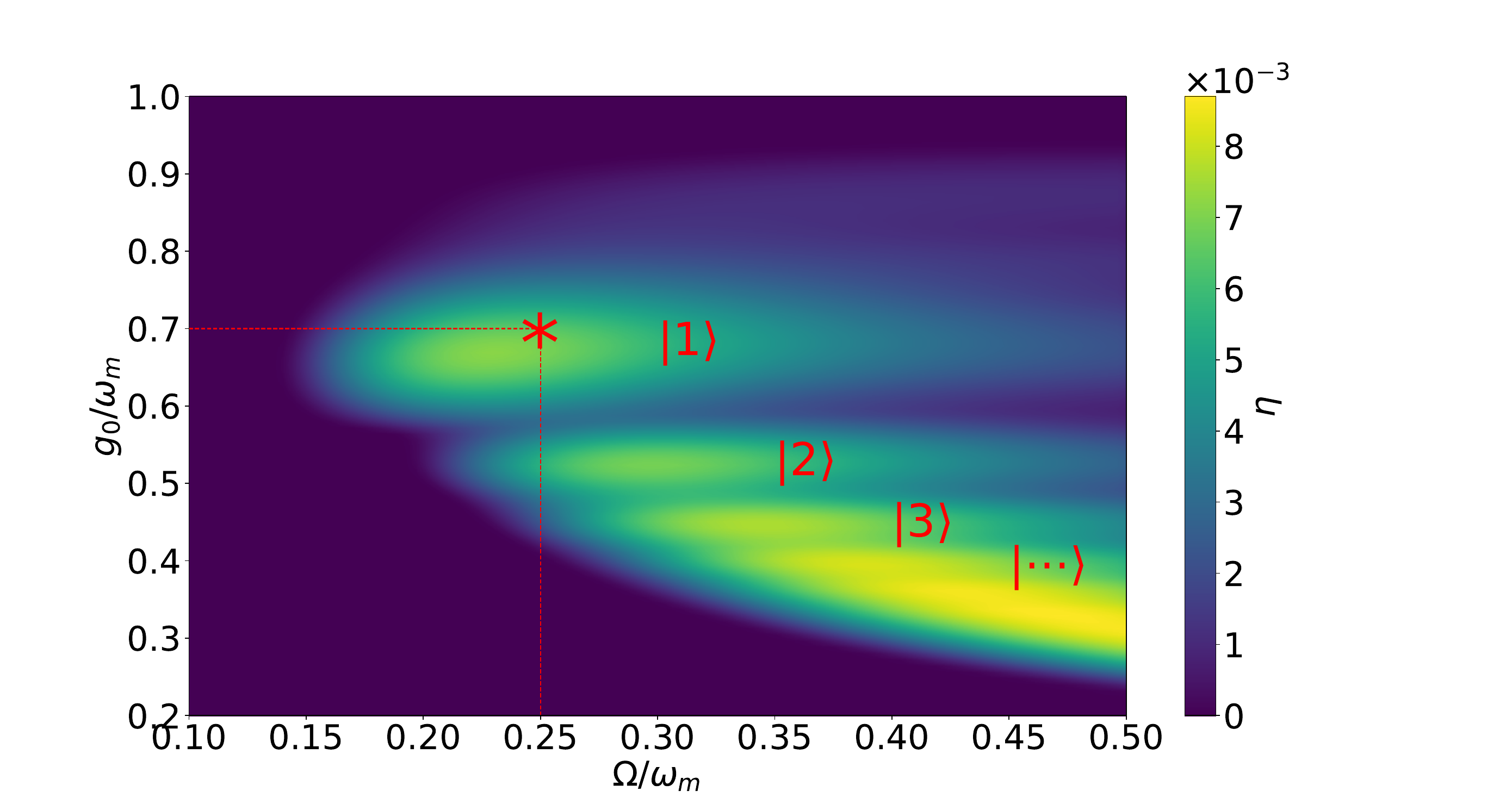}
\caption{Negativity factor $\eta$ when varying the coupling constant $g_0$ and the Rabi frequency $\Omega$, for $\epsilon = 0.3\ \omega_m$. The notation $\ket{n}$ indicates the appearance of a sharp peak for $n$ in $P_n$, and the star represents the parameters of Fig.~\ref{Tomography}. The other parameters, in unit of $\omega_m$, are: $\Gamma = 0.01$, $\gamma = \gamma_\phi = 10^{-4} $.}
\label{Eta}
\end{figure}

\textit{Beyond the secular approximation.} – 
The secular approximation fails when the condition \(\epsilon \ll \Gamma, \gamma\) is no longer valid, but we can still numerically solve Eq. (\ref{Dissip_diag}) for the density matrix \(\rho_s\) \cite{Qtip}.  
As before, we find large Fano factors at the mechanical transition and nonclassical steady states for strong coupling \(g_0\) (see Fig.~S2 in the SM), confirming that our previous interpretations remain valid even in the presence of nonvanishing coherences.
\textit{Cavity optomechanics.} -
The method presented and the behavior that we found allows to shed a new light on the cavity optomechanics in the regime $g_O / \omega_m \lesssim1$, where $g_O$ is the single photon optomechanical coupling, which has been intensively investigated in the past \cite{Kippenberg_Cavity_2007,Nation_Nonclassical_2013,aspelmeyer_cavity_2014,lorch_laser_2014,qian_quantum_2012,Wise_Nonclassical_2024}.
Actually, the large $g_O$ coupling generates a Kerr term in the Hamiltonian of the form of 
$n_c^2 g_O^2/\omega_m$, where $n_c$ is the number of photons in the cavity.
The anharmonicity allows to single out two states, as in the superconducing qubits \cite{Kosh_charge_2007} or more recently mechanical qubits \cite{pistolesi_proposal_2021,yang_mechanical_2024}.
Thus, it is possible to blue-detune the laser to a transition involving only two cavity states, say $n_c = 0,1$, while the others are detuned.
We find that the cavity optomechanical model can then be mapped onto the Hamiltonian (\ref{H_s}) for a TLS coupled to an oscillator via $\sigma_z$ (see SM).
This induces the simplifications that we exploited to obtain the results presented above, which thus apply to both systems.
%

\textit{Conclusion.} -
In this paper, we investigate the properties of a mechanical oscillator strongly coupled to a TLS. 
When the TLS is weakly driven by a coherent source tuned to the first blue sideband, we find that increasing the drive intensity causes the oscillator to transition from a thermal to a self-oscillation regime. 
This transition is signaled by large photon fluctuations, which we found to be strongly correlated with the phonon fluctuations through the relation (\ref{Bruit}).
At the mechanical transition, the photon statistics are characterized by a large photon Fano factor, which can be measured.
As the electro-mechanical coupling is increased, we observe the emergence of non-classical steady mechanical states. 
These states are associated with sharp phonon distributions that resemble Fock states. 
We then propose a method to extract the Wigner distribution from the experimental photon-emission spectrum.
This method utilizes the diagonal form of the density matrix in the secular approximation, making it a broadly applicable approach.
Therefore, we expect it to extend beyond the specific system studied here.
We finally showed that this approach applies to cavity optomechanical systems in similar regimes, recovering and extending previous results obtained through more complex numerical methods.

\textit{Acknowledgement.} -
The authors acknowledge financial support from the French {\em Agence Nationale de la Recherche} through contract ANR IMOON ANR-22-CE47-0015, and from the French government in the framework of the University of Bordeaux’s France 2030 program / GPR LIGHT.

\bibliographystyle{apsrev4-1}
\bibliography{Article.bib}

\end{document}


\title{Two-Level System Nanomechanics in the Blue-Detuned Regime}

\author{Guillaume Bertel}
\author{Clément Dutreix}
\author{Fabio Pistolesi}

\affiliation{Université de Bordeaux, CNRS, LOMA, UMR 5798, F-33400 Talence, France}

\date{\today}
        
\maketitle
\onecolumngrid
\appendix

\newcommand{\fabio}[1]{{\color{blue}#1}}
\renewcommand{\figurename}{Fig.}
\renewcommand{\thefigure}{S\arabic{figure}}


\section{Diagonalization of the Hamiltonian}
We recall the full system Hamiltonian including the fictitious cavity representing the laser drive in the initial basis, 
\begin{equation}
H_{\rm tot} = \omega_m a^\dagger a + \omega_L b^\dagger b + \frac{\omega_0}{2} \sigma_z + g_0 (a + a^\dagger) \sigma_z + \frac{\Omega}{2} \left( \sigma_+ b + \sigma_- b^\dagger \right) \,,
\end{equation}
where $a$ ($b$) is the annhiliation operator for the phonon (photon) in the oscillator (the cavity), and $\sigma_i$ is the $i$-Pauli operator.
We first apply the Lang-Firsov (LF) transformation $U_{1} = e^{-g_0(a^\dagger-a)\sigma_z/\omega_m}$ to diagonalize exactly the electro-mechanical coupling.
In the LF basis, the Hamiltonian reads
\begin{equation}
\label{Ham_LF}
H_1= U^\dagger_1 H_{\rm tot} U_1 = \omega_m a^\dagger a + \omega_L b^\dagger b + \frac{\omega_0}{2} \sigma_z + \frac{\Omega}{2} \left[ \sigma_+ b e^{2g_0(a^\dagger-a)/\omega_m} + \sigma_- b^\dagger e^{-2g_0(a^\dagger-a)/\omega_m} \right] - \frac{g_0^2}{\omega_m} \mathbb{I} \,.
\end{equation}
It is clear from this Hamiltonian that any change of the state of the TLS (via the $\sigma_\pm$ operators) is associated to a displacement of the oscillator (via the $e^{\pm 2g_0(a^\dagger-a)/\omega_m}$ operators).
The last term of equation (\ref{Ham_LF}) is the so-called polaronic energy, and it can be discarded when studying the equation of motion.
We then diagonalize the Hamiltonian keeping the source interaction term by applying standard degenerate perturbation theory.
For this we consider a frequency of the laser $\omega_L = \omega_0 + \omega_m + \epsilon$, where $|\epsilon| \ll \omega_m$ is the detuning from the first blue-detuned sideband.
Therefore the Hamiltonian (\ref{Ham_LF}) is approximated to be bloc-diagonal, with coupling terms only between the states $\ket{e,n,N}$ and $\ket{g,n-1,N+1}$.
Here $\ket{i,n,N}$ are the eigenstates of the Hamiltonian for $\Omega=0$ as indicated in the main text.
We thus obtain the eigenstates in presence of drive (this defines the unitary transformation $U$)
\begin{equation}
\label{eigenstates}
\begin{split}
& \ket{+,n,N} = \sqrt{\frac{\Omega_n - \epsilon}{2\Omega_n}} \ket{e,n,N} + \sqrt{\frac{\epsilon + \Omega_n}{2\Omega_n}} \ket{g,n-1,N+1} \,, \\
& \ket{-,n,N} = \sqrt{\frac{\epsilon + \Omega_n}{2\Omega_n}}  \ket{e,n,N} - \sqrt{\frac{\Omega_n-\epsilon}{2\Omega_n}}  \ket{g,n-1,N+1} \,,
\end{split}
\end{equation}
where we introduced the Rabi splitting $\Omega_n \equiv \sqrt{(\Omega W_{n,n-1})^2 + \epsilon^2}$, with $W_{n,m}=\braket{n|e^{2g_0(a^\dagger-a)/\omega_m}|m}$ the Franck-Condon factor.
Let us introduce $\ket{\mu}$ with $\mu=\pm$ 
so that 
\begin{equation}
\label{eigenstates}
\ket{\mu,n,N} = \alpha_\mu(n) \ket{e,n,N} + \beta_\mu (n) \ket{g,n-1,N+1} \,.
\end{equation}
The associated eigenvalues read
\begin{equation}
E_{\pm,n,N} = n\omega_m + N\omega_L + \frac{\omega_0}{2}  + \frac{1}{2} \left( \epsilon \pm \Omega_n \right) \,.
\end{equation}
We finally obtain the diagonal Hamiltonian
\begin{equation}
\label{Hamilt}
H_2 = U^\dag H_1 U =
\sum_n \frac{\Omega_n}{2} \ket{n}\bra{n} \otimes \sigma_z + \omega_m a^\dagger a + \omega_L b^\dagger b + \left(\frac{\omega_0 + \epsilon}{2} \right) \mathbb{I} \ ,
\end{equation}
where now the Pauli matrix is in the space of the eigenstates ($\pm$) and the last term can be discarded.
%


\section{Master Equation }
We now take into account the coupling to the environment and derive a master equation for the 
system density matrix in the basis where the Hamiltonian is diagonal.
In the initial basis, the dissipation is described via the interaction Hamiltonians:
\begin{equation}
\begin{split}
& H_{\rm EM} = (\sigma_+ + \sigma_-) \otimes E_{1} \ , \\
& H_{\rm mec} = (a+a^\dagger) \otimes E_{2} \ .
\end{split}
\end{equation}
Where $E_{1}$ and $E_{2}$ are operators of the electromagnetic and mechanical environments.
In the LF basis, these Hamiltonians become:
\begin{equation}
\label{H_env_lf}
\begin{split}
& U_1^\dagger H_{\rm EM} U_1  = \left( \sigma_+ e^{2g_0(a^\dagger-a)/\omega_m}  + \sigma_- e^{- 2g_0(a^\dagger-a)/\omega_m} \right) \otimes E_{1} \ ,\\
& U_1^\dagger H_{\rm mec} U_1 = \left( a+a^\dagger-2\frac{g_0}{\omega_m} \sigma_z \right) \otimes E_{2} \ .
\end{split}
\end{equation}
Within the Born and Markov approximations, the master equation of the system reads \cite{schlosshauer_decoherence_2008}
\begin{equation}
\label{BM}
\begin{split}
\frac{d}{dt}\rho_S^{(I)}(t) & = - \int_0^\infty d\tau \ \sum_\alpha \left\{C_\alpha(\tau) \left[S_\alpha(t) S_\alpha(t-\tau) \rho_S^{(I)}(t) - S_\alpha(t-\tau) \rho_S^{(I)}(t) S_\alpha(t) \right] \right. \\
& \left. + C_\alpha(-\tau) \left[\rho_S^{(I)}(t) S_\alpha(t-\tau) S_\alpha(t) - S_\alpha(t) \rho_S^{(I)}(t) S_\alpha(t-\tau) \right] \right\} 
=
\left[{\cal L}_m + {\cal L}_o\right]
\rho_S^{(I)}(t) 
\ .
\end{split}
\end{equation}
Here, $\rho_S^{(I)}(t)$ is the density matrix of the system in the interaction picture and $S_\alpha(t)$ indicates the operators of the system involved in the interactions with the environment having a correlation function $C_\alpha(t) =\braket{E_\alpha(t) E_\alpha(0)}$.
The master equations naturally splits into two parts, one induced by the coupling to the mechanical environment (${\cal L}_m $) and one to that of the optical environment (${\cal L}_o$).
%
%
Therefore, from Eqs. (\ref{H_env_lf}) we have $S_\alpha = U^\dagger x_z(a+a^\dagger-2g_0\sigma_z/\omega_m)U$ for the dissipation induced by the mechanical environment; and $S_\alpha = U^\dagger (\sigma_+ e^{2g_0(a^\dagger-a)/\omega_m} + \sigma_- e^{-2g_0(a^\dagger-a)/\omega_m})U$ for the dissipation induced by the electromagnetic environment.
In the following, we detail the calculations of one of the terms of ${\cal L}_m $:
\begin{equation}
\label{BM_1}
{\rm{BM}}_1 =  \int_0^\infty d\tau \ \sum_\alpha C_\alpha(\tau)  S_\alpha(t-\tau) \rho_S^{(I)}(t) S_\alpha(t) \ .
\end{equation}
The time-evolution of the $S_\alpha(t)$ operators is obtained by writing explicitly 
\begin{equation}
S_\alpha(t) = \sum_{i,j} S_{ij} \ e^{i \omega_{ij} t} \ket{i} \bra{j} \ ,
\end{equation}
where $\ket{i}$ and $\ket{j}$ are the eigenstates of the system in the diagonal basis. 
Using the general notation in Eq.~(\ref{eigenstates}) and calling $\ket{i} = \ket{\mu,n,N}$ and $\ket{j} = \ket{\mu',n',N'}$, we find for instance for the operator $U^\dagger a U$
\begin{equation}
(U^\dagger a U)_{ij} = \left[ \alpha_\mu(n) \alpha_{\mu'}(n') \sqrt{n'-1} + \beta_\mu(n) \beta_{\mu'}(n') \sqrt{n'-2} \right] \delta_{n,n'-1} \delta_{N,N'} \ ,
\end{equation}
where $\delta_{n,n'}$ is the Kronecker symbol. 
We define $\mu = \{-1,1\}$ such that $\sigma_z \ket{\mu} = \mu \ket{\mu}$.
It follows 
\begin{equation}
\begin{split}
(U^\dagger a U)(t) = & e^{-i \omega_m t} \sum_{\mu,\mu',n} \left[ \alpha_\mu(n) \alpha_{\mu'}(n+1) \sqrt{n} + \beta_\mu(n) \beta_{\mu'}(n+1) \sqrt{n-1} \right] \\
& e^{-i(\mu \Omega_n - \mu' \Omega_{n+1})t/2} \ket{\mu,n}\bra{\mu',n+1} \ .
\end{split}
\end{equation}
Considering that $\Omega_n \ll \omega_m$, at lowest order the time-evolution of $(U^\dagger a U)(t)$ is given by $e^{-i \omega_m t}$.
A similar derivation for $(U^\dagger a^\dagger U)(t)$ leads to the dominant time dependent term $e^{i \omega_m t}$.
Finally we find for $(U^\dagger \sigma_z U)(t)$ 
\begin{equation}
(U^\dagger \sigma_z U)(t) = \sum_{\mu,\mu',n} \left[ \alpha_\mu(n) \alpha_{\mu'}(n) - \beta_\mu(n) \beta_{\mu'}(n) \right] e^{i(\mu - \mu')\Omega_n t/2} \ket{\mu,n}\bra{\mu',n} \ .
\end{equation}
We see that the time dependence of $(U^\dagger \sigma_z U)(t)$ is controlled by the low frequency $\Omega_n$.
In Eq.~(\ref{BM_1}), we have combinations of these operators (oscillating as $e^{\pm i \omega_m t}$) and of $\rho_S^{(I)}(t)$ (that depends on time over the low frequency scale given by $\gamma$ and $\Gamma$). 
Since we assumed the resolved sidebands limit $\gamma, \Gamma \ll \omega_m$, we can use the rotating wave approximation and neglect the fast oscillating terms.
Therefore Eq.~(\ref{BM_1}) reduces to
\begin{equation}
\label{term_1}
\begin{split}
\rm{BM}_1 = \int_0^\infty d\tau \ C_\alpha(\tau) &  \left[ a(t-\tau) \rho_S^{(I)}(t) a^\dagger(t) + a^\dagger(t-\tau) \rho_S^{(I)}(t) a(t) \right. \\
& \left. + \left( 2\frac{g_0}{\omega_m} \right)^2 \sigma_z(t-\tau) \rho_S^{(I)}(t) \sigma_z(t) \right]    \ .
\end{split}
\end{equation}
Now we perform the integral. It is known from the Sokhotski–Plemelj theorem \cite{Breuer_Theory_2007} that
\begin{equation}
\label{Int}
\int_0^\infty d\tau \ C(\tau) e^{\pm i \kappa \tau} = -\frac{i}{2 \pi} \int_{-\infty}^\infty d\omega \ \tilde{C}(\omega) \mathcal{P}\left( \frac{1}{\omega \mp \kappa} \right) + \frac{1}{2}\tilde{C}(\pm \kappa) \ .
\end{equation}
On that equation, $\tilde{C}(\omega)$ is the Fourier transform of $C(t)$, while $\mathcal{P}$ indicates the Cauchy principal value.
For the following we always neglect this last term, as it only induces a shift of the energy of the system.
Therefore Eq.~(\ref{BM_1}) reads
\begin{equation}
\label{term_2}
\begin{split}
\rm{BM}_1 & = \frac{1}{2} \left[\tilde{C}(\omega_m) a(t) \rho_S^{(I)}(t) a^\dagger(t) + \tilde{C} (-\omega_m) a^\dagger(t) \rho_S^{(I)}(t) a(t)\right]  \\
& + \left( 2\frac{g_0}{\omega_m} \right)^2 \int_0^\infty d\tau \ C(\tau) \sigma_z(t-\tau) \rho_S^{(I)}(t) \sigma_z(t) \ .
\end{split}
\end{equation}
In order to perform the remaining integral in Eq.~(\ref{term_2}), we consider that the spectrum of correlation of the mechanical environment is flat around $\omega \sim 0$.
This means $\tilde{C}(\Omega_n) \approx \tilde{C}(0)$, for any $\Omega_n$.
We finally obtain
\begin{equation}
\rm{BM}_1 = \frac{1}{2} \left[ \tilde{C}(\omega_m) a(t) \rho_S^{(I)}(t) a^\dagger(t) + \tilde{C} (-\omega_m) a^\dagger(t) \rho_S^{(I)}(t) a(t) + \left( 2 \frac{g_0}{\omega_m} \right)^2 \tilde{C}(0) \sigma_z(t) \rho_S^{(I)}(t) \sigma_z(t) \right] \ .
\end{equation}
We define the oscillator damping rate $(1+\gamma n_B) = \tilde{C}(\omega_m)$ and the pure dephasing rate $\gamma_\phi = \left( 2 \frac{g_0}{\omega_m} \right)^2 \tilde{C}(0)$.
%
%
Repeating this derivation for all the other terms in the BM equation (\ref{BM}), we finally obtain in the Schrödinger picture Eq.~(2) from the main text:
\begin{align}
\label{Dissip_diag}
\frac{d}{dt} \rho_s  = & -i [H_1, \rho_s] + \gamma (n_B + 1) \mathcal{D}(a)\rho_s + \gamma n_B \mathcal{D}(a^\dagger) \rho_s  \notag  \\
&  + \Gamma \mathcal{D} \left( \sigma_- e^{-\frac{g_0}{\omega_m}(a^\dagger-a)} \right) \rho_s + \gamma_\phi \mathcal{D}(\sigma_z) \rho_s  \ .
\end{align}
%


%
\section{Rate equation}
In the regime where the secular approximation applies, the master equation reduces to a Pauli rate equation.
Defining the reduced probabilities $P_{\mu,n} = \sum_N P_{\mu,n,N}$ and using Eqs.~(5) and (6) from the main text, we obtain
\begin{equation}
\begin{split}
    \dot{P}_{\mu,n} & =  - P_{\mu,n} \sum_{\mu'} \left[ \sum_{n'} \Gamma_{\mu,n \to \mu',n'} + \gamma_{\mu,n \to \mu',n+1} + \gamma_{\mu,n \to \mu',n-1} \right] \\
     & + \sum_{\mu'} \left[ \sum_{n'} P_{\mu',n'} \Gamma_{\mu',n' \to \mu,n} + P_{\mu',n+1} \gamma_{\mu',n+1 \to \mu,n} + P_{\mu',n-1} \gamma_{\mu',n-1 \to \mu,n} \right] \ .
\end{split}
\end{equation}
Then, the probability to have $n$ phonon in the system reads  
\begin{equation}
    P_{n} = \sum_{\mu} \left[ P_{\mu,n} \alpha_\mu^2(n) + P_{\mu,n+1} \beta_{\mu}^2(n+1) \right] \ ,
\end{equation}
with $\alpha_\mu(n)$ and $\beta_\mu(n)$ the coefficient defined in Appendix~(A).

\section{Effect of an pure intrinsic dephasing rate}
\begin{figure}
    \centering
    \includegraphics[scale=0.32]{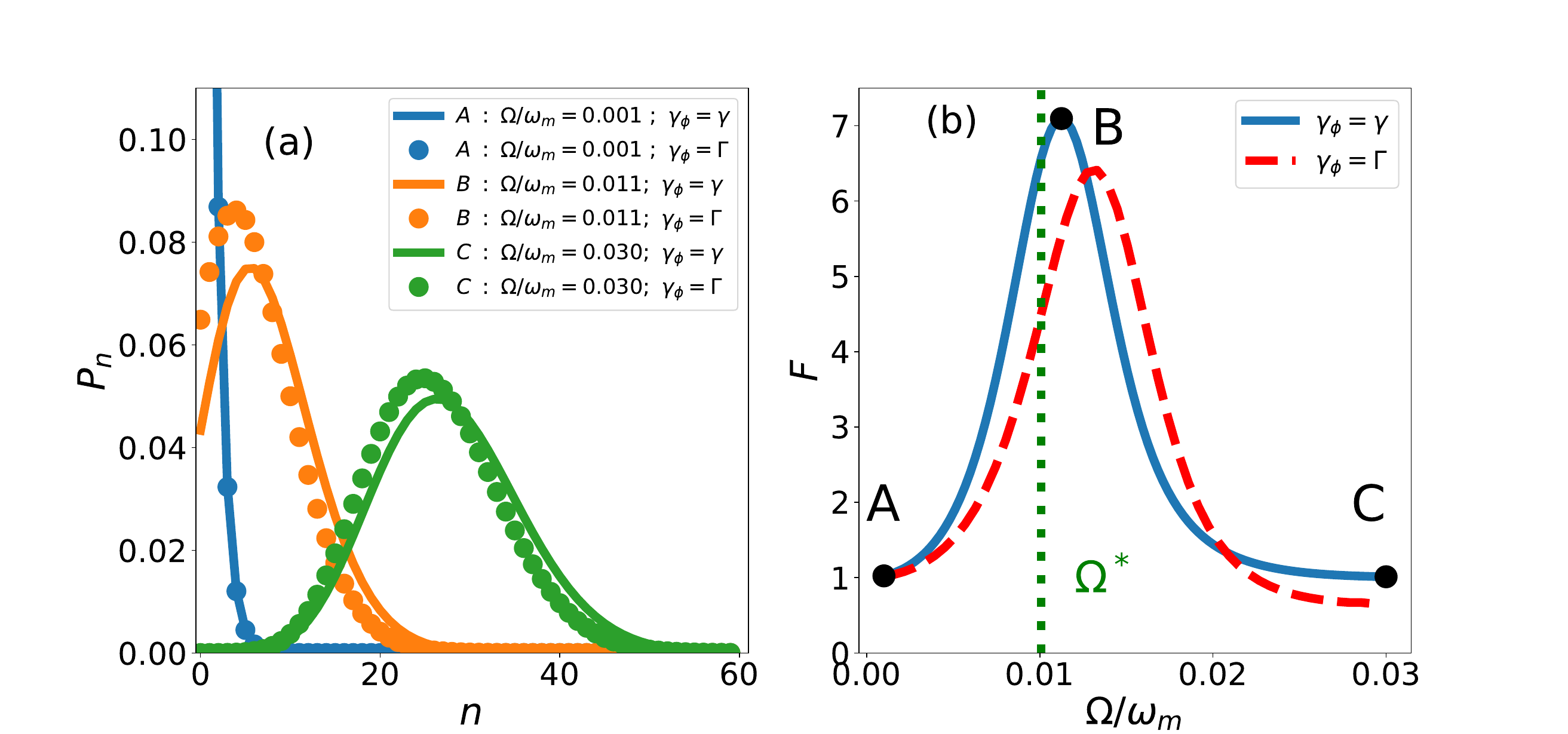}
    \includegraphics[scale=0.3]{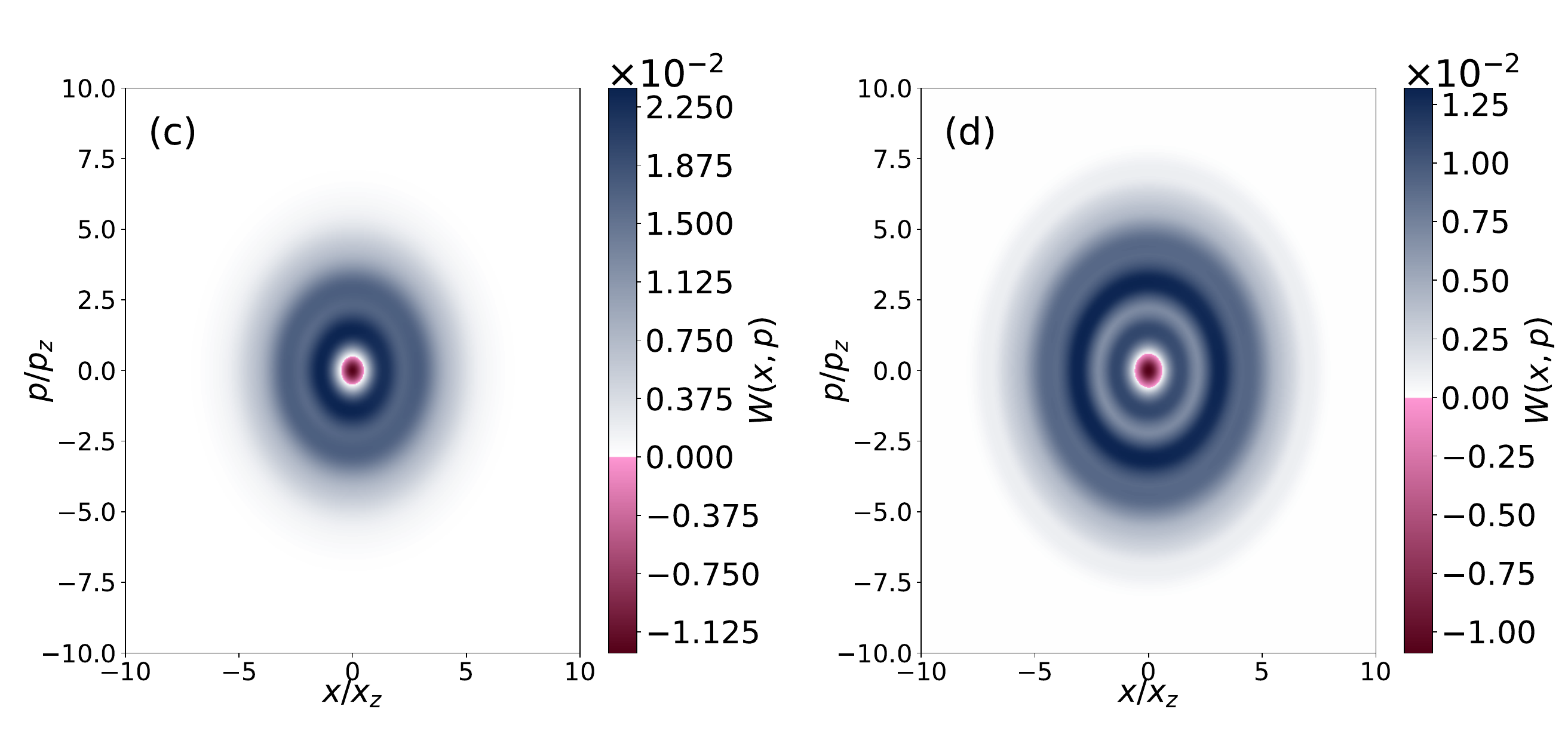}
    \caption{(a) Phonon probability distribution $P_n$ for three different driving intensities labeled as $A$, $B$, and $C$, alternatively for $\gamma_\phi = \gamma$ (solid lines) and $\gamma_\phi = \Gamma$ (dots). 
    (b) Fano factor of the light emitted by the TLS, for $\gamma_\phi = \gamma$ (blue solid line) or $\gamma_\phi = \Gamma$ (red dashed line), as a function of the drive intensity $\Omega/\omega_m$. In both pictures, the parameters in units of $\omega_m$ are: $g_0 = 0.1$, $\Gamma = 0.01$, $\gamma = 10^{-4} $, $\epsilon = 0.01$.
    (c) Wigner distribution for $\gamma_\phi = \gamma$ and (d) Wigner distribution for $\gamma_\phi = \Gamma$, for the parameters in units of $\omega_m$: $g_0 = 0.7$, $\Gamma = 0.01$, $\gamma = 10^{-4} $, $\Omega = 0.25$, $\epsilon = 0.3$.} 
    \label{Gamma_phi}
\end{figure}

In the main text, we mention that introducing an intrinsic pure dephasing rate $\Gamma_\phi$ for the TLS is equivalent to increasing $\gamma_\phi$, and specify that this does not significantly affect the results presented.
Figure~(\ref{Gamma_phi})(a,b) compares the phonon probability distributions and the Fano factor in two scenarios: without intrinsic TLS dephasing (as in the main text) and with an intrinsic dephasing rate of  $\gamma_\phi=\Gamma$. All other parameters remain unchanged from Fig. (2) of the main text. The close agreement between the curves demonstrates that even a large intrinsic pure dephasing term $\gamma_\phi$ has minimal effect. 
Similarly, Fig~(\ref{Gamma_phi})(c,d) presents the Wigner distribution for the parameters of Fig. 3 in the main text, with $\gamma_\phi = \gamma$ in (c) and $\gamma_\phi = \Gamma$ in (d).
Once again, these results highlight the weak impact of the pure dephasing rate on the system's behavior.

\section{Symmetry in the photon flux}
The flux of photons emitted by the TLS associated to the creation (or annihilation) of $p$ phonons in the oscillator is defined as
\begin{equation}
    I^{(p)} = \sum_{\mu,\eta,n} P_{\mu,n} \Gamma_{\mu,n \to \eta,n+p} \ .
\end{equation}
From Eq.~(5) from the main text, the transition rates read
\begin{equation}
\Gamma_{\mu,n \to \eta,n+p} = \Gamma \alpha^2_\mu(n) \beta^2_\eta(n+p) W^2_{n,n+p-1} \ .
\end{equation}
Using this last equation and the relation between the coefficients $\alpha$ and $\beta$, the expression for $I^{(p)}$ reduces to
\begin{equation}
I^{(p)} = \Gamma \sum_{\mu,n} P_{\mu,n} \alpha^2_\mu(n) W^2_{n,n+p-1} \ .
\end{equation}
We then explicit the Fanck-Condon factors, using the following equation from \cite{koch_theory_2006} :
\begin{equation}
W_{n,m} = {\rm{sgn}}(M-N)^{N-M} \sqrt{\frac{N!}{M!}} \left(2\frac{g_0}{\omega_m}\right)^{M-N} e^{-2(g_0/\omega_m)^2} \mathcal{L}_{N}^{(M-N)} \left[ \left(2\frac{g_0}{\omega_m} \right)^2 \right] \ ,
\end{equation}
where $M=\rm{max}(m,n)$, $N=\rm{min}(m,n)$ and $\mathcal{L}_{n}^{(\alpha)}$ is the generalized Laguerre polynomial of order $n$.
Expanding this expression at the lowest order in $g_0/\omega_m$, one can readily obtain for $p \in \mathbb{N}$ :
\begin{equation}
\begin{split}
& W_{n,n+p} \sim \sqrt{\frac{n!}{(n+p)!}} \left(2\frac{g_0}{\omega_m}\right)^p \frac{(n+p)!}{p!n!} \ , \\
& W_{n,n-p}\sim \sqrt{\frac{(n-p)!}{n!}} \left(2\frac{g_0}{\omega_m}\right)^p \frac{n!}{p!(n-p)!}\ .
\end{split}
\end{equation}
The symmetry relation given in the main text $I^{(1+p)} \simeq I^{(1-p)}$ is verified as long as $W_{n,n+p}^2 / W_{n,n-p}^2 \simeq 1$.
This implies that
\begin{equation}
\frac{(n+p)!(n-p)!}{(n!)^2} \simeq 1 \ ,
\end{equation}
which is true for $n$ large or $p$ small. 
In the opposite limit, when $p$ approaches $n$, the relation above breaks down.
However, larger $p$ implies smaller values of the Franck-Condon factor, and consequently of $I^{(p)}$. 
In general, only the terms for $0 \le p \le 2 $ really contribute to the total flux.
%


\section{Relation between the photon flux and the mean number of phonon}
From the Fermi golden rule, one can write the mean value of the photon flux as
\begin{equation}
\bar{I} = \sum_{\mu,\eta,n,m} P_{\mu,n} \Gamma |\braket{\eta,m| \sigma_- e^{-g_0(a^\dagger-a)/\omega_m} |\mu,n}|^2 \ .
\end{equation}
Using the closure relation and considering that
the lone state $|e,0\rangle$ that does not couple to the other states has a vanishing occupation ($P_{e,0} \simeq 0$), one obtains:
\begin{equation}
\bar{I} = \Gamma \sum_{\mu,n} P_{\mu,n} \alpha_\mu(n)^2 \ .
\end{equation}
An expansion of the transition rates at the lowest order in $g_0/\omega_m$ shows that $\Gamma_{-,n\to+,m} \sim 1$, while $\Gamma_{+,n\to -,m} \sim \mathcal{O}(g_0/\omega_m)^4$ and $\Gamma_{\pm,n\to \pm,m} \sim \mathcal{O}(g_0/\omega_m)^2$.
Therefore the dynamics is separated  between a fast and a slow part.
In the stationary regime, one can consider only the slow dynamics, which implies that $P_{-,n} \simeq 0$.
Additionally, the coefficient $\alpha_+$ can be expanded as 
\begin{equation}
\alpha_+^2(n) \simeq \left( \frac{\Omega g_0}{\epsilon \omega_m} \right)^2 n \equiv A n \ .
\end{equation}
This leads to the photon flux 
\begin{equation}
\bar{I} = \Gamma A \sum_n P_{+,n} n \ .
\end{equation}
On the other side, the mean number of phonons in the oscillator reads
\begin{equation}
\bar{n} = \sum_{\mu}\sum_{n=1}^\infty  n \left[ P_{\mu,n} \alpha^2_\mu(n) + P_{\mu,n+1}\beta^2_\mu(n+1)  \right] \ .
\end{equation}
In the framework of the approximations previously described, this reduces to 
\begin{equation}
\bar{n}  = (A+1) \sum_{n}  P_{+,n}  n - 1 \simeq \sum_{n}  P_{+,n}  n - 1 \ ,
\end{equation}
where in the last step we used $A\ll 1$.
We finally obtain the equation given in the main text
\begin{equation}
\label{I_vs_n_bar}
 \bar{I} = \Gamma A (\bar{n} + 1) \ .  
\end{equation}
We note that this equation is valid for small electro-mechanical coupling. 
Additionally, the approximation $P_{-,n} \approx 0$ remains valid only for $\Omega<\Omega^*$,
{\em i.e.} for values below the threshold for the self-oscillation. 
\begin{figure}
    \centering
    \includegraphics[scale=0.35]{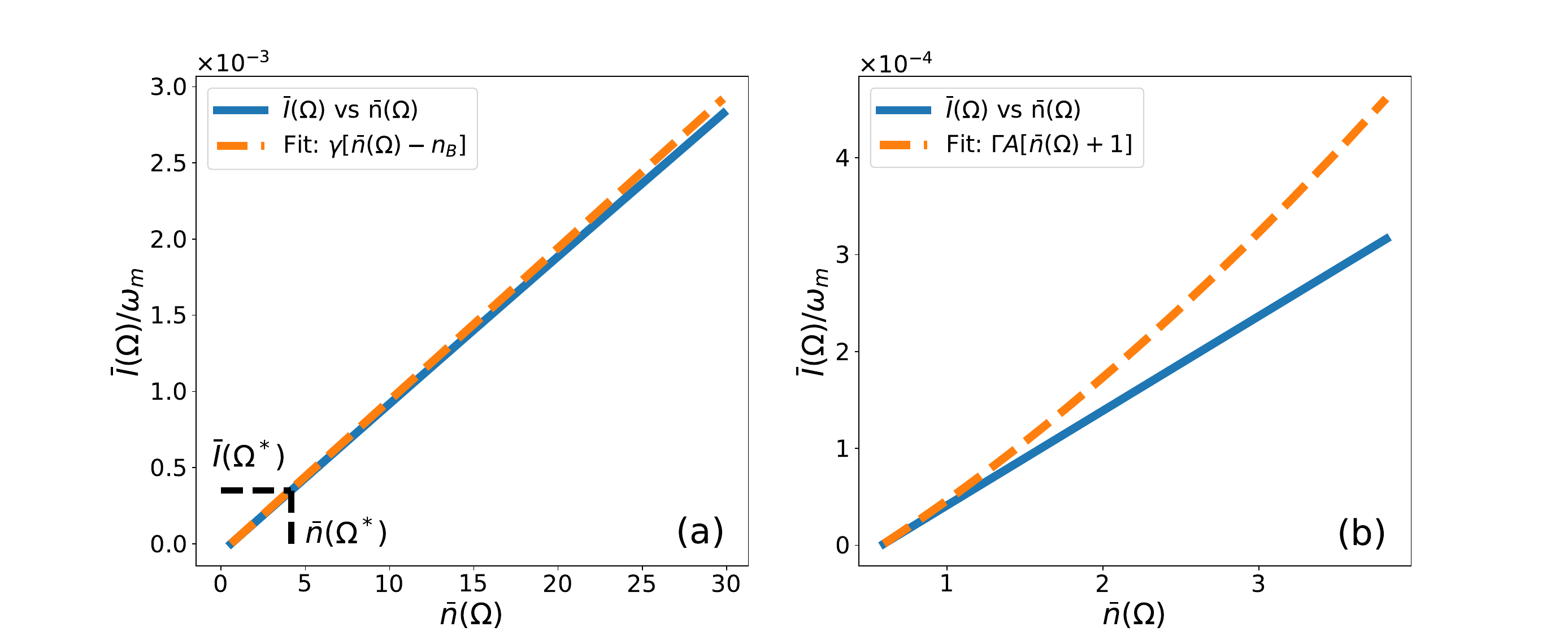}
    \caption{Mean photon flux and mean phonon number plotted as parameterized curves of $\Omega$: $\left(\bar{I}(\Omega),\bar{n}(\Omega)
    \right)$. (a) Comparison between the numerical data and Eq.~(8) from the main text. (b) Comparison between the numerical data and Eq.~(\ref{I_vs_n_bar}), for $\Omega < \Omega^*$. The parameters are the same as for the Fig. 2 in the main text.}
    \label{I_vs_n}
\end{figure}
The comparison between the numerical values of $\bar{I}$ as a function of $\bar{n}$ and the Eq.~ (\ref{I_vs_n_bar}) is displayed of Fig.~\ref{I_vs_n}(b). 
On this figure we only plotted the values of $\bar{I}$ and $\bar{n}$ for $\Omega < \Omega^*$, as for larger laser intensity the equation is no longer valid.


\section{Numerical method for the noise spectrum of the phonon number}
To determine numerically the noise spectrum of the phonons in the oscillator, we follow standard methods used to determine the spectrum of operators when the system is described by the evolution of a probability ($P$) of occupation of given states (see for instance \cite{pistolesi_self-consistent_2008}).
One has the evolution equation for the vector $P$ as $\dot{P} = \hat M P$.
The matrix $\hat M$ admits different left and right eigenvectors, with complex eigenvalues $\lambda_i$ such that ${\rm Re}\lambda_i < 0$,
apart for $\lambda_0=0$.
One then introduces operators acting in the space
of probability so that the operator average can be obtained by a scalar product 
$(w_0,\hat O \ v_0)$, where 
$w_0$ and $v_0$ are the left and right, respectively, eigenstates with vanishing eigenvalue of $\hat M$.
%
%
Let us introduce the variation of the number $n$ as 
$\delta \hat {n} = \hat n - \bar n$.
One can then calculate the classical correlation function as follows:
\begin{equation}
S_{\rm nn}(t) = (w_0, \delta \hat{n} \ e^{\hat Mt} \delta \hat{n}\ v_0)  .
\end{equation}
Considering that $S_{nn}(t)$ is symmetric in time, the Fourier transform reads
\begin{equation}
S_{\rm nn}(\omega) =  \int_0^\infty dt \  (w_0, \delta \hat {n} \ e^{\hat Mt} \delta \hat {n} \  v_0) (e^{i\omega t} + e^{-i \omega t}) \ .
\end{equation}
By integration we obtain  
\begin{equation}
S_{\rm nn}(\omega) = - 2 \left( w_0, \delta \hat {n} \ \frac{\hat M}{\hat M^2+\omega^2} \delta \hat {n} \  v_0  \right) \ .
\end{equation}
The noise spectrum for the phonon at zero frequency finally reads $S_{\rm nn}( 0) = \lim\limits_{\omega \to 0} S_{\rm nn}( \omega) $.


\section{Full counting statistics}
In the main text, we introduced the generating function as $e^{{\cal S}(\chi,t)} = \sum_{i,k} P_{i,k}(t) e^{ik \chi} = \braket{e^{ik \chi}}$, where $P_{i,k}(t)$ is the probability to be in the state $\ket{i}$ at time $t$ while $k$ photons are in the environment. We follow for the calculation Ref. \cite{Bagrets_full_2003} (see also \cite{pistolesi_full_2004}).
From this we retrieve all the cumulants of the number of photons emitted in the EM environment.
Since we are interested by the statistics of the stationary photon flux, we take the time-derivative of these cumulants in the steady state $(t \to \infty)$.
That way we obtain
\begin{equation}
\begin{split}
\label{i_bar_and_sigma_i}
    & \bar{I} = \frac{\partial \dot{\cal S}_{t \to \infty}(\chi,t)}{\partial(i\chi)}|_{\chi \to 0} \ , \\
    & S_{II} = \frac{\partial^2 \dot{\cal S}_{t \to \infty}(\chi,t)}{\partial(i\chi)^2}|_{\chi \to 0} \ .
\end{split}
\end{equation}

To derive the generating function, we define the Fourier-like transform $P_i(\chi,t) = \sum_k P_{i,k} (t) e^{i k \chi}$ such that $e^{{\cal S}(\chi,t)} = \sum_i P_i(\chi,t)$, and we introduce the vector $P(\chi,t)$ whose elements are the $P_i(\chi,t)$.
Then, in a similar way as in the Appendix G one can write a rate equation 
\begin{equation}
\label{rate_eq}
    \dot{P}(\chi,t) = \hat M(\chi) P(\chi,t) \ .
\end{equation}
Here $\hat M(\chi)$ is the rate matrix where all the non-diagonal elements related to transitions induced by the EM environment are multiplied by $e^{i \chi}$.
Like in the former section, $\hat M(\chi)$ admits right eigenvectors $V_n (\chi)$ with complex eigenvalues $\lambda_n(\chi)$ such that ${\rm Re}\lambda_n < 0$.
Solving Eq.~(\ref{rate_eq}) in the basis of $V_n (\chi)$ leads to
\begin{equation}
P(\chi,t) = \sum_n \alpha_n(\chi) V_n (\chi) e^{\lambda_n (\chi) t} \ ,
\end{equation}
with $\alpha_n(\chi)$ some irrelevant constant number.
For $t \to \infty$, all the elements of this sum are exponentially damped.
The only elements that remains is the one for which ${\rm Re}\lambda_n (\chi)$ is the greatest, that we label $\lambda_0 (\chi)$.
It follows
\begin{equation}
P_{t \to \infty}(\chi,t)  \simeq \alpha_0(\chi) V_0 (\chi) e^{\lambda_0 (\chi) t} \ .
\end{equation}
The generating function in the steady state then reads 
\begin{equation}
{\cal S}_{t \to \infty}(\chi,t) = ln \left[ \sum_i \alpha_0(\chi) V_0^{(i)} (\chi) \right] + \lambda_0(\chi) t \ .
\end{equation}
The time-derivative reduces to
\begin{equation}
\dot{\cal S}_{t \to \infty}(\chi,t) = \lambda_0(\chi) \ .
\end{equation}
Finally, from Eqs. (\ref{i_bar_and_sigma_i}) we obtain
\begin{equation}
     \bar{I} = \left.\frac{\partial \lambda_0(\chi)}{\partial(i\chi)}\right|_{\chi \to 0} \ , \qquad 
     S_{II} = \left.\frac{\partial^2 \lambda_0(\chi)}{\partial(i\chi)^2}\right|_{\chi \to 0} \ .
\end{equation}


%
\section{Reconstruction of the density matrix}

In the main text, we discuss how the full density matrix (and hence the non-classical state) can be measured directly from the photon-emission spectrum of the system.
In this section we give few additional details.

The photon emission spectrum is defined as
\begin{equation}
S_{\rm ph}(\omega) 
= 
\int_{-\infty}^{+\infty} dt \braket{\sigma_+^d(t)\sigma_-^d(0)} e^{i \omega t} 
=
2 {\rm Re} \left[\int_{0}^{+\infty} dt \  \braket{\sigma_+^d(t)\sigma_-^d(0)} e^{i \omega t} \right]\ ,
\end{equation}
where we introduced the Pauli operator in the diagonal basis $\sigma_\pm^d = U^\dagger \sigma_\pm e^{\pm 2 g_0 (a^\dagger - a)/\omega_m} U$, and made use of the relation 
\begin{equation}
    \braket{\sigma_+^d(t)\sigma_-^d(0)}^*=
    \braket{\sigma_+^d(-t)\sigma_-^d(0)}
\end{equation}
to relate the negative time correlation function to the positive one.
In the secular approximation the master equation 
$d{\rho}_s /dt = {\mathcal{L}}{\rho}_s$ has a stationary solution that is diagonal in the basis of the eigenstates ($\ket{\sigma}$) of the Hamiltonian:
$\rho_{\rm st} = \sum_\sigma P_\sigma \ket{\sigma}\bra{\sigma}$.
Using the quantum regression theorem we can then write the photon-emission spectrum as
\begin{equation}
S_{\rm ph}(\omega) = 2 {\rm Re} \left[ \sum_\sigma P_\sigma \int_{0}^{+\infty} dt \  {\rm Tr} \left( {\sigma}_+^d (0) e^{{\mathcal{L}}t } {\sigma}_-^d \ket{\sigma}\bra{\sigma} \right) e^{i\omega t} \right] \ .
\end{equation}
Defining $\lambda_i$, $\ket{v_i}$, and $\bra{w_i}$, the eigenvalue, the right, and left eigenvectors of ${\mathcal{L}}$, respectively, the equation becomes: 
\begin{equation}
S_{\rm ph}(\omega) = 2 {\rm Re} \left[  \sum_{i,\sigma} P_\sigma \ {\rm Tr} \left( {\sigma}_+^d \ket{v_i}  \bra{w_i} {\sigma}_-^d \ket{\sigma}\bra{\sigma} \right) \int_{0}^{+\infty} dt \ e^{(\lambda_i + i\omega) t } \right] \ .
\end{equation}
We denote $\lambda_0=0$ the eigenvalue corresponding to the the stationary solution $|v_0\rangle$ and $\bra{w_0}$ the corresponding left eigenvalue. One can show that 
the trace operation is equivalent to the projection onto the vector $\bra{w_0}$.

The term for $i=0$ does not contribute to the sum since 
$\langle \sigma_+\rangle=0$.
For the stability of the system ${\rm Re}(\lambda_i) <0$ for all $i>0$, leading to a converging integral that can be performed:
\begin{equation}
S_{\rm ph}(\omega) = 
\sum_\sigma P_\sigma f_\sigma(\omega) \ .
\label{AnaSpectrum}
\end{equation}
where
\begin{equation}
f_\sigma(\omega) = 
- 2 {\rm Re} \left[  \sum_{i}  \bra{y_0} \check{\sigma}_+^d \ket{v_i}  \frac{\bra{w_i} \check{\sigma}_-^d \ket{\sigma}}{(\lambda_i + i\omega)} \right] .
\end{equation}
The expression (\ref{AnaSpectrum}) obtained is perfectly suited to be used in a fit procedure to extract $P_\sigma$ from the experimental spectrum.

In order to show the feasibility of the procedure we investigate the efficiency of the method proposed. 
Let us assume that we have measured the emission spectrum on a set of frequencies $\{ \omega_i \}$ for i=1, \dots , $N$, covering the non-vanishing spectrum region. 
We thus have a set of measured data $\{ S^{\rm (ex)}_{\rm ph}(\omega_i) \}$.
We can define the $\chi^2$ function as
\begin{equation}
\label{Chi_squared}
    \chi^2 = \sum_{i=1}^N 
    \left[ {S^{\rm (ex)}_{\rm ph}(\omega_i) - \sum_\sigma P_\sigma f_\sigma(\omega_i) \over \Delta S_i} \right]^2 \ ,
\end{equation}  
where $\Delta S_i$ are the estimated error on the data. 
In the following, for simplicity, we assume 
that $\Delta S_i=\Delta S$ independent on $i$.
We now proceed to the minimization of the $\chi^2$ with the constraint that the distribution is normalized. 
We thus eliminate $P_0$ from the minimization with the condition $P_0 = 1-\sum_{\sigma \neq 0} P_\sigma$.
We then look for the saddle point given by the condition
 $\partial \chi^2 / \partial P_{\sigma \neq 0} = 0$, which in matrix form reads: $A P = B$, with
\begin{equation}
\begin{split}
& A_{\sigma\sigma'} = \sum_i \frac{1}{\Delta_i^2} \left[f_\sigma(\omega_i)-f_0(\omega_i) \right] \left[f_{\sigma'}(\omega_i)-f_0(\omega_i) \right] \ , \\
& B_\sigma = \sum_i \frac{1}{\Delta_i^2} \left[f_\sigma(\omega_i)-f_0(\omega_i) \right] \left[S^{\rm (ex)}_{\rm ph}(\omega_i)-f_0(\omega_i) \right] \ .
\end{split}
\end{equation}  
The error drops from the minimization equation.
The linear equation can be solved numerically leading to the solution $ P = A^{-1}B$.

To test the procedure, we simulate the expected experimental spectrum by first computing the exact spectrum from the full master equation. 
Then, we add a random noise to each point, following a Gaussian distribution of mean $S_{\rm ph} (\omega_i)$ and of standard deviation $p\times \max\limits_{\omega} [S_{\rm ph}(\omega)]$, with $p$ a parameter such that $0<p<1$.

\begin{figure}[H]
    \centering
    \includegraphics[scale=0.19, trim = 2cm 0cm 4.5cm 2cm, clip]{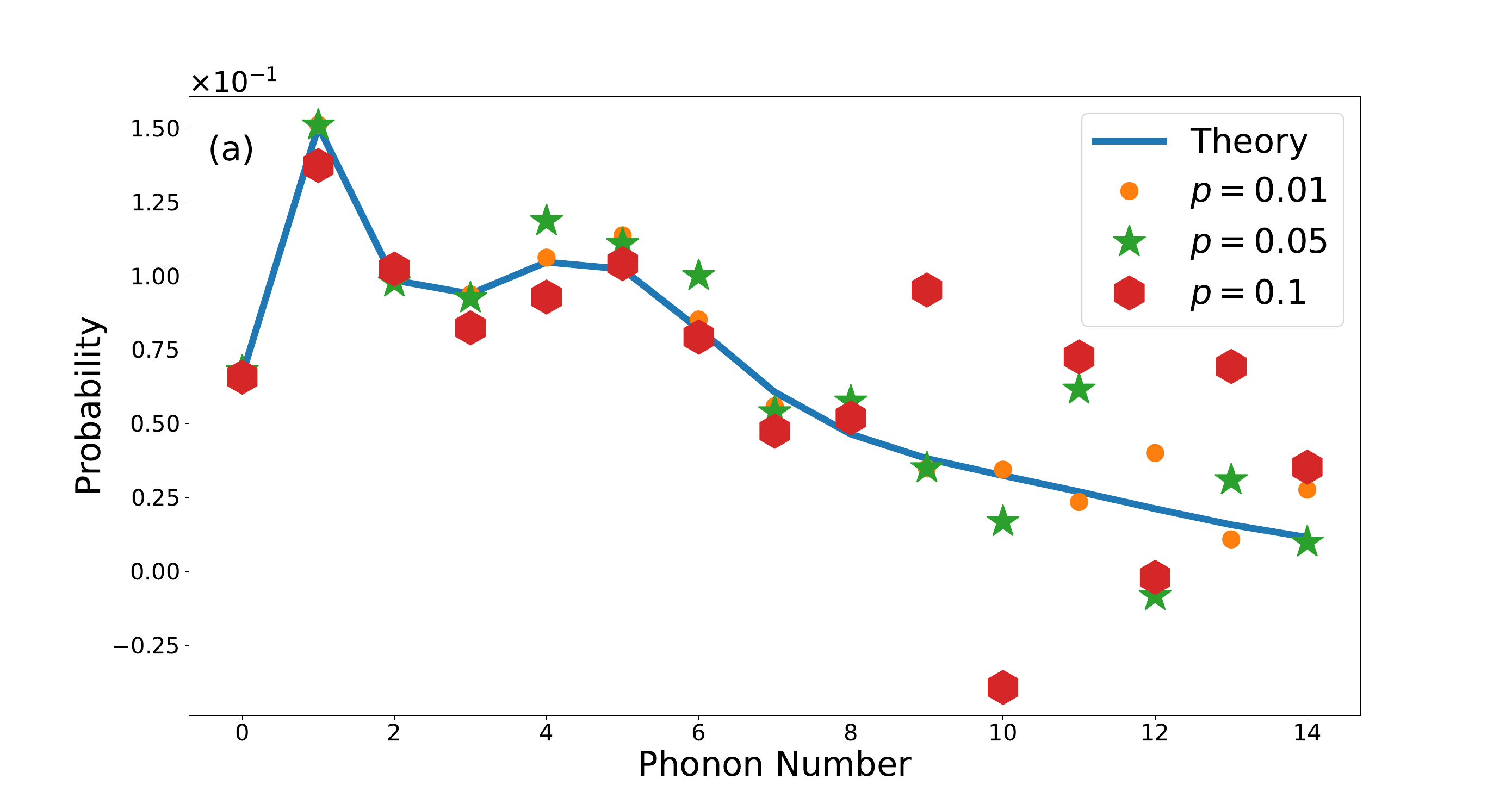}
    \includegraphics[scale=0.19, trim = 1.75cm 0cm 4.5cm 2cm, clip]{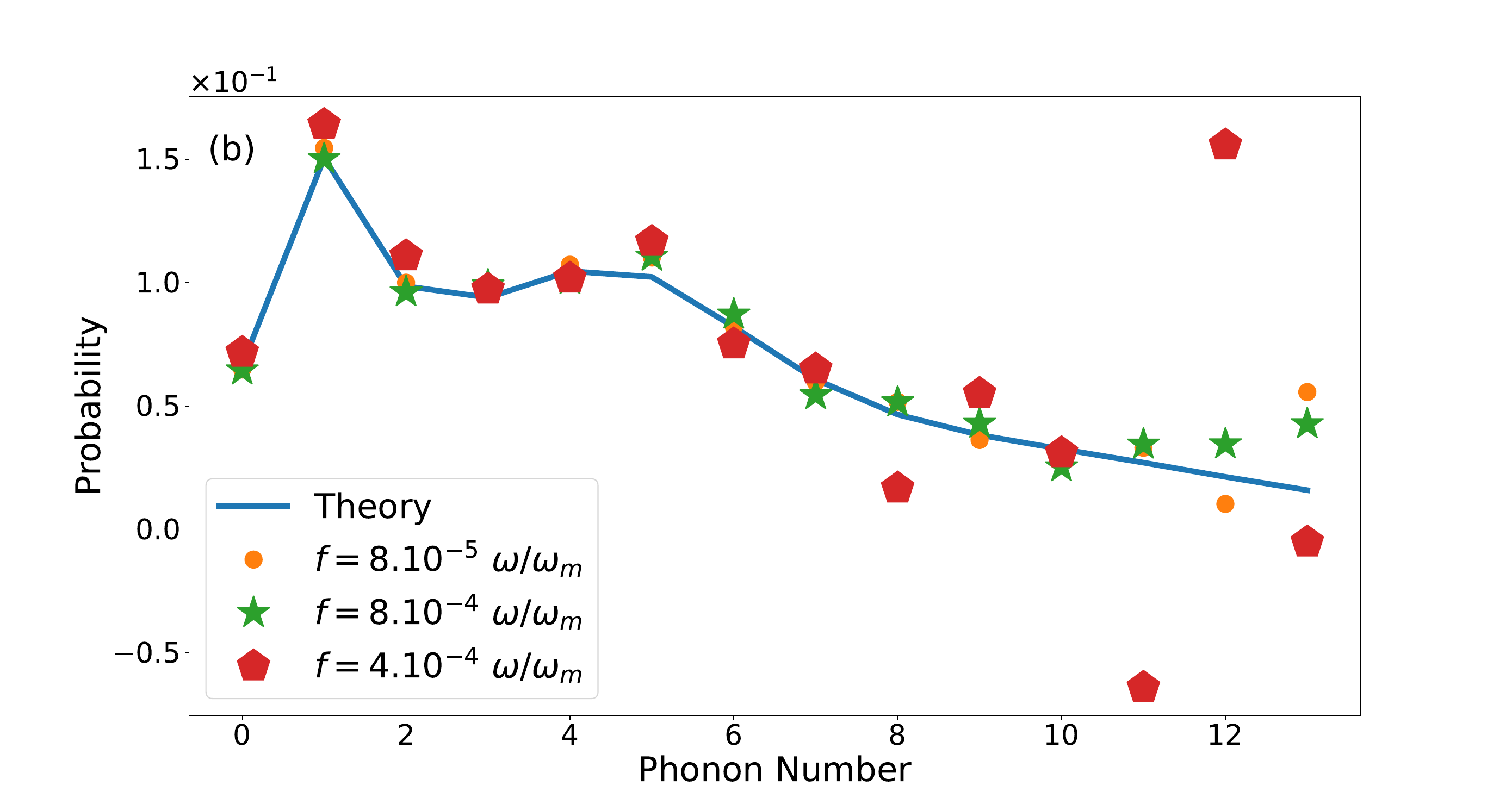}
    \caption{Phonon probability distribution $P_n$ predicted by the simulation (blue solid line), compared with the $P_n$ obtained via the fitting procedure. 
    In (a) the introduced noise error is $p=0.01$ (orange dots), $p=0.05$ (green stars) or $p=0.1$ (red hexagons), and the sampling rate is $f=(\omega_{i+1}-\omega_i) / \omega_m=8.10^{-5}$.
    In (c) the sampling rate is $f=8.10^{-5}$ (orange dots), $f=8.10^{-4}$ (green stars) or $f=4.10^{-4}$ (red hexagons), and the percentage of error is $p=0.01$.
    The other parameters are the same as in Fig~3(b) of the main text : $g_0 = 0.7$, $\Omega = 0.25$, $\Gamma = 0.01$, $\gamma = \gamma_\phi = 10^{-4}$, and $\epsilon = 0.3$.}
    \label{Noise_and_sampling}
\end{figure}

The result of the phonon probability distribution fitted with the method presented and for different values of the noise intensity $p$ is shown in Fig~\ref{Noise_and_sampling}(a) [for examples of noisy spectra see Fig~\ref{Error_bar}(e,f,g)].
This Figure shows that even for $p=0.1$, the method allows to retrieve remarkably well the predicted probability distribution.
For this value, the reduced $\chi_r^2=\chi^2/N$ reads $\chi^2_r = 0.9987$, indicating that the agreement between the simulated experimental spectrum and the theoretical model is consistent with the expected error variance. 
Figure~\ref{Noise_and_sampling}(b) shows instead the dependence on the sampling frequency spacing.

In Fig~\ref{Noise_and_sampling}(b), we set the noise to be $1\% $ of the maximum of the signal and we vary the sampling frequency spacing $f=(\omega_{i+1}-\omega_i) / \omega_m$.
As expected, a lower sampling rate induces more errors on the $P_n$, as several pics in the emission-spectrum can no longer be distinguished.

We discuss now the expected error for each value of the $P_n$ predicted by the minimization procedure.
To estimate the error, we identify the range in which varying a single parameter, while still minimizing all the others, causes $\chi^2$ to remain within a $95 \%$ confidence interval.
This approach provides insight into whether certain values of $P_n$ are more sensitive to fluctuations than others.
In order to calculate these quantities we begin by rewriting Eq.~(\ref{Chi_squared}) as 
\begin{equation}
    \chi^2 = P^T A P + 2B P + C \ ,
\end{equation}
with $C = \sum_i \left\{ S^{\rm (ex)}_{\rm ph}(\omega_i)^2 \left[ 1 - 2 f_0(\omega_i) \right] + f_0(\omega_i)^2  \right\}$, and $A$ and $B$ the matrices defined above.
We then expand $\chi^2$ to the second order around its minimum $P^{(0)}$ : $\chi^2(P) =\chi^2(P^{(0)}) + \delta P^T A  \delta P$, with $\delta P = P - P^{(0)}$. 
Next, we express the variation $\delta\chi^2=\chi^2(P) -\chi^2(P^{(0)})$ as a function of a single element of $\delta P$, say $\delta P_k$:
\begin{equation}
\label{delta_chi_2}
\begin{split}
    \delta \chi^2 & = A_{kk}\delta P_k^2 + 2 \delta P_k \sum_{i\neq k}A_{ik} \delta P_i + \sum_{i\neq k , j\neq k}A_{ij} \delta P_i \delta P_j \\
    & \equiv  A_{kk}\delta P_k^2 + 2 \delta P_k (B')^T \delta P' + (\delta P')^T A' \delta P' \ .
    \end{split}
\end{equation}
Here, we introduced $\delta P'$ the vector $\delta P$ with its $k$-th row removed, $A'$ the matrix $A$ with its $k$-th row and $k$-th column removed, and $B'$ the $k$-th column of $A$ with its $k$-th row removed.
To account for the potential correlations between $P_k$ and the other elements $P_{i\neq k}$, we determine the values of $\delta P_{i\neq k}$ that minimize $\delta \chi^2$ for the chosen value of $\delta P_k$.
To achieve this, we look for the saddle point given by the condition $\partial (\delta \chi^2)/ \partial (\delta P_{i\neq k}) = 0$, which reads in the matrix form
\begin{equation}
    A' \delta P' = - \delta P_k B' \ .
\end{equation}
Inverting this relation and injecting it in the Eq~(\ref{delta_chi_2}), it follows
\begin{equation}
    \delta \chi^2 =  \left\{ A_{kk} - 2 (B')^T (A')^{-1} B' + [(A')^{-1}B']^T A' [(A')^{-1}B']\right\} \delta P_k^2 \ ,
\end{equation}
or alternatively 
\begin{equation}
\label{delta_chi_2_2}
    \delta P_k  =  \sqrt{\delta \chi^2} \left\{ A_{kk} - 2 (B')^T (A')^{-1} B' + [(A')^{-1}B']^T A' [(A')^{-1}B']\right\}^{-1/2}  \ .
\end{equation}
Equation~(\ref{delta_chi_2_2}) relates the fluctuation $\delta P_k$ to a given variation $\delta \chi^2$, while accounting for the reorganization of all other $\delta P_{i \neq k}$.
To define the error bars $\delta P_k$, we select for $\delta\chi^2$ the value $\delta \chi^2_{\rm max} = \chi^2_{\rm min} - \chi^2_{\rm max}$, where $\chi^2_{\rm min}$ is the minimum found by the minimization procedure, and $\chi^2_{\rm max}$ is determined with the condition
\begin{equation}
\int_{-\infty}^{\chi^2_{\rm max}} d\chi^2 f(\chi^2) = 0.95 \ .
\end{equation}
Here $f(\chi^2)$ is the probability distribution of $\chi^2$.
As the number of degrees of freedom $\nu$ is quite large ($\nu \sim 10^{5}$), this is well approximated by a normal distribution of mean $\nu$ and of standard deviation $\sqrt{2\nu}$.
This particular value of $\delta \chi^2_{\rm max}$ ensures that $\delta P_k$ lies within a $95 \%$ confidence interval.
Figure~[\ref{Error_bar}(a,b,c)] compares the predicted phonon probability distribution to the distribution obtained using the fitting method for a sampling rate $f=8.10^{-5}$, including the error bars computed with Eq.~(\ref{delta_chi_2_2}).
The error bars grow significantly with increasing noise intensity, highlighting the strong correlation between the $P_k$.
Naturally, the error becomes more pronounced for larger $k$.
Figure~[\ref{Error_bar}(d,e,f)] shows the noisy emission spectra obtained for the values of noise intensity used in Fig.~[\ref{Error_bar}(a,b,c)].

\begin{figure}[H]
    \centering
    \includegraphics[scale=0.3, trim = 2cm 0cm 4.5cm 1.5cm, clip]{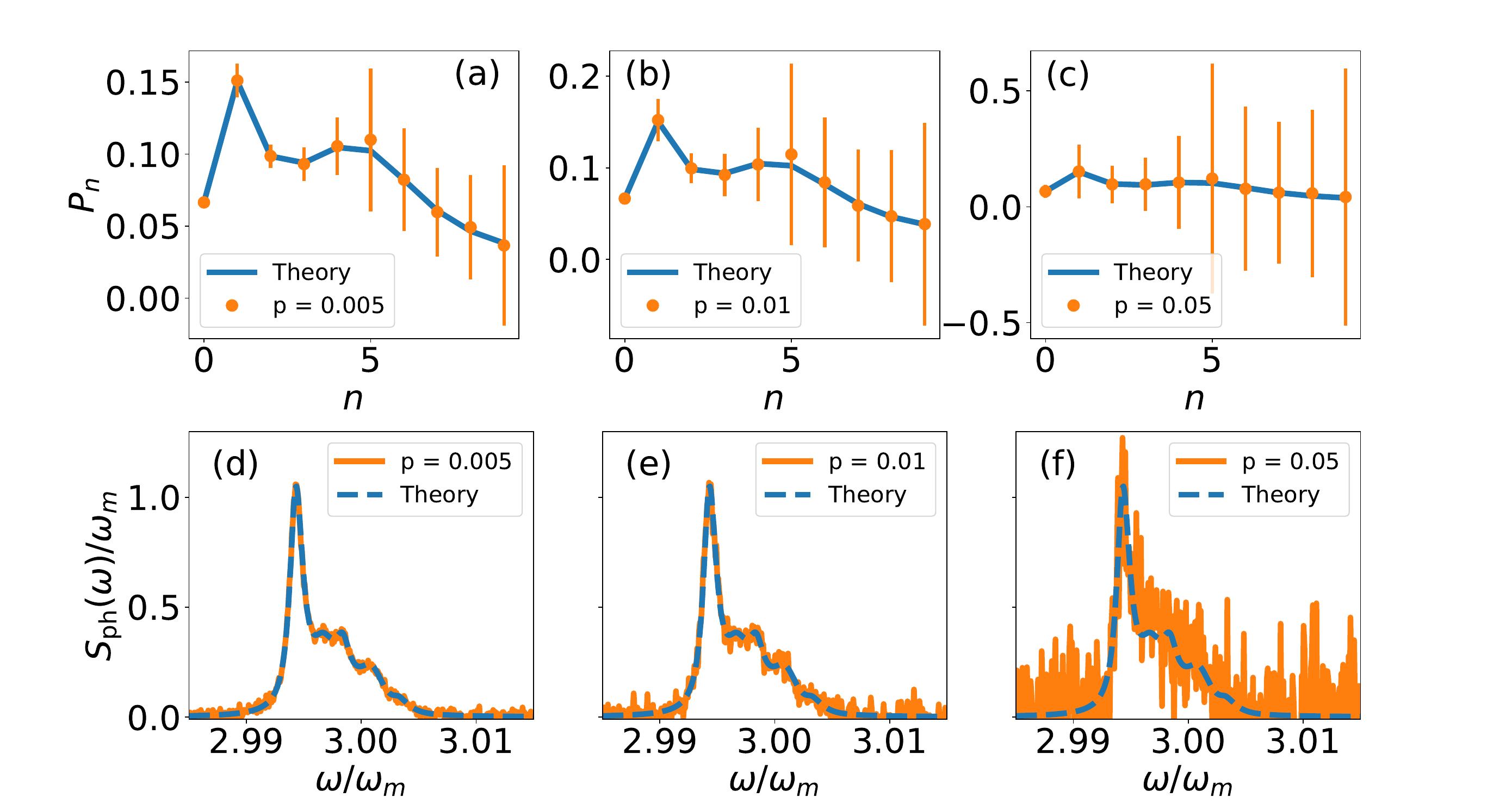}
    \caption{(a,b,c) Phonon probability distribution $P_n$ predicted by the simulation (blue solid line) and by the minimization method (orange dots), with the error-bars obtained via Eq.~(\ref{delta_chi_2_2}) for $\delta \chi^2_{\rm max}$, for a noise parameter $p=0.05$ (a), $p=0.01$ (b) and $p=0.5$ (c).
    (d,e,f) Photon-emission spectrum obtained by the resolution of the master equation (blue dashed line), compared to the noisy spectrum with a noise parameter $p=0.05$ (d), $p=0.01$ (e) and $p=0.5$ (f).
    The other parameters are the same as in Fig~3(b) of the main text : $g_0 = 0.7$, $\Omega = 0.25$, $\Gamma = 0.01$, $\gamma = \gamma_\phi = 10^{-4}$, and $\epsilon = 0.3$.}
    \label{Error_bar}
\end{figure}

%
%
%
%
%

\rem{Figure~(\ref{Error_bar}) shows the distribution $P_n$ with the error-bars obtained from Eq.~(\ref{error}), for the sampling rate $f=8.10^{-5}$ and the noise intensity $p=0.1$, considering a variation $\delta\chi^2$ as large as $\chi^2(P^{(0)})$.
This corresponds to assuming a fluctuation of the difference of the data with the theory of the order of the error-bars.
The result of the Figure indicates that the statistical error on the probabilities $P_n$ is relatively small, and that the error is uniform.
WHAT ABOUT THE FACT THAT ACTUALLY SOMETHING WAS NOT UNIFORM \textcolor{red}{$\rightarrow$ It is actually not uniform, but the values of the error bars are too similar to be distinguishable of the Figure.}

\begin{figure}[H]
    \centering
    \includegraphics[scale=0.25, trim = 2cm 0cm 4.5cm 2cm, clip]{SM_Pn_error_bar_delta_chi_10.pdf}
    \caption{Phonon probability distribution $P_n$ predicted by the simulation (blue solid line), with the error-bars obtained via Eq.~(\ref{error}), for $\delta \chi^2 = \chi^2(P^{(0)})$.
    The other parameters are the same as in Fig~3(b) of the main text : $g_0 = 0.7$, $\Omega = 0.25$, $\Gamma = 0.01$, $\gamma = \gamma_\phi = 10^{-4}$, and $\epsilon = 0.3$.}
    \label{Error_bar}
\end{figure}}


\section{Beyond the secular approximation}

\begin{figure}[H]
    \centering
    \includegraphics[scale=0.35]{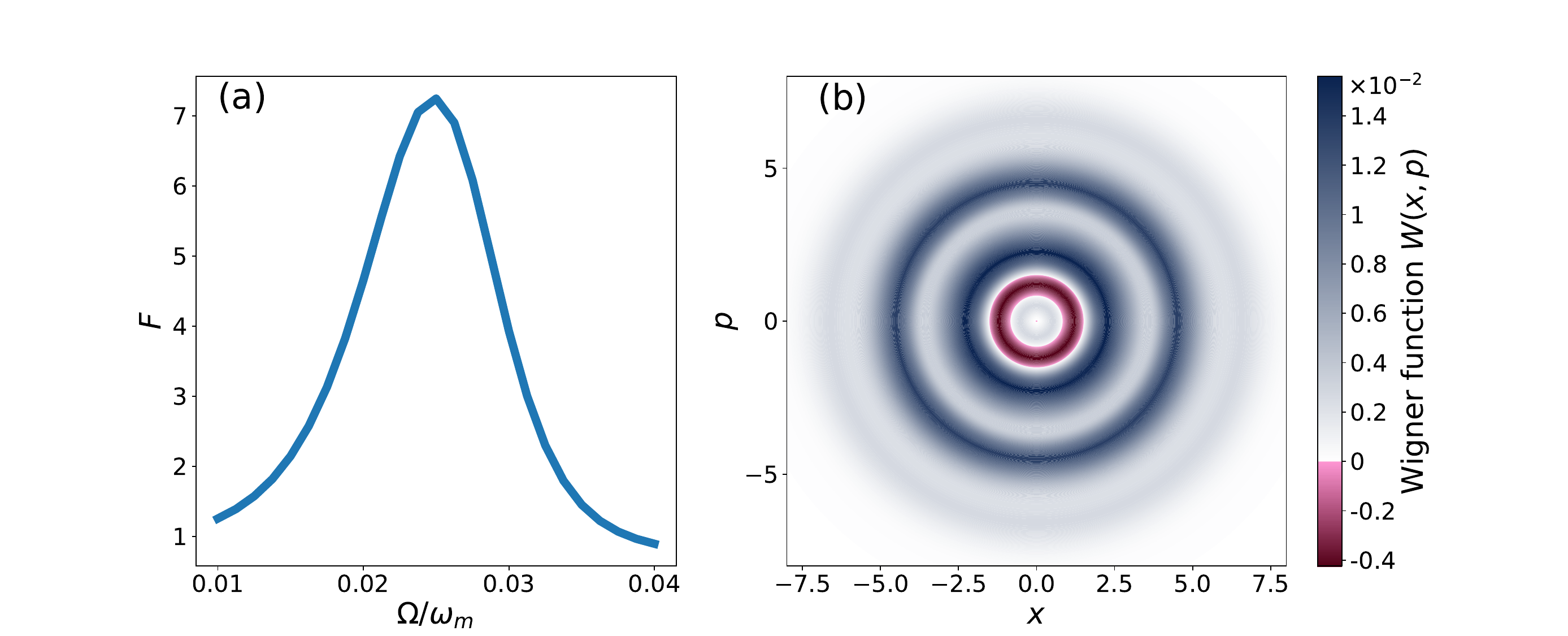}
    \caption{(a) Fano factor of the light emitted by the TLS, obtained by integration of the second order correlation function of the photons $g^{(2)}(t)$, for a coupling constant $g_0 = 0.1 \omega_m$. (b) Wigner distribution of the phonon for $g_0 = 0.5 \omega_m$ and $\Omega = 0.05 \omega_m$, with negativity $\eta = 1.3 \%$. The parameters in units of $\omega_m$ are : $\Gamma = 0.1$, $\gamma = 10^{-4}$, $\epsilon = 0.05$, $k_B T = 1$. }
    \label{non_sec}
\end{figure}

For $\epsilon < \Gamma$, the secular approximation breaks down.
However, as specified in the main text, the master equation can still be solved numerically.
It is then possible to recover the results discussed previously, such as the large Fano factor at the mechanical transition [see Fig~\ref{non_sec}(a)], or the non-classical mechanical state [see Fig~\ref{non_sec}(b)].
On these latter Figures, $\epsilon = \Gamma /2$ implies that the secular approximation does not apply.
In this case the full counting statistic method described on Appendix F cannot be used to derive the Fano factor, so we use instead the Mandel equation \cite{mandel_sub-poissonian_1979}
\begin{equation}
    F = 1 + 2\bar{I} \int_0^\infty dt \ \left[ g^{(2)}(t) - 1 \right] \ ,
\end{equation}
where $g^{(2)}(t)$ is the second order correlation function of the photons emitted by the TLS, in the stationnary regime.

\section{Application of our model to opto-mechanical cavities}
\begin{figure}
    \centering
    \includegraphics[scale=0.35]{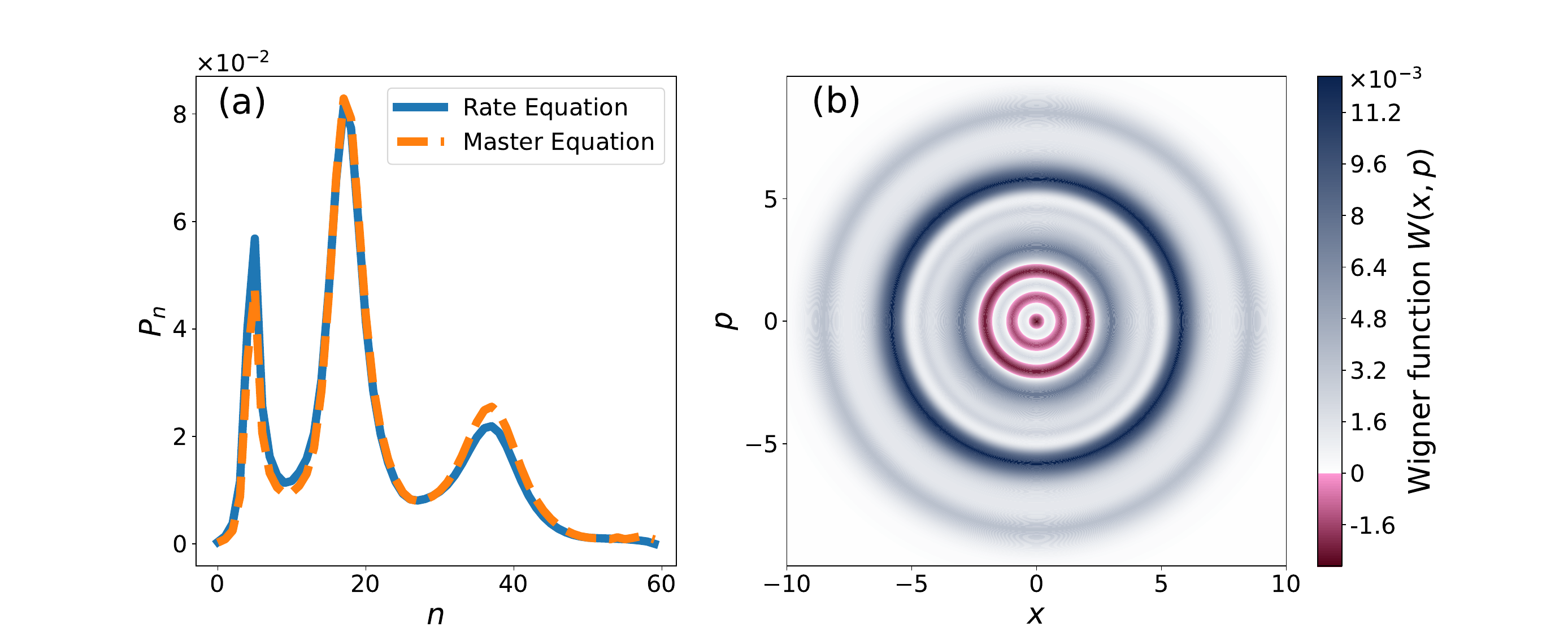}
    \caption{(a) Phonon probability distribution in the oscillator in the steady state, solving the Pauli rate equation (blue solid line) or the full master equation (dashed orange line). (b) Wigner function associated, with negativity $\eta = 1.42 \%$. The parameters in units of $\omega_m$ are : $g_O = 0.8$, $\Omega = 0.1$, $\kappa = 0.01$, $\gamma = 10^{-4}$, $\epsilon = 0.05$, $T = \hbar \omega_m/k_B$. }
    \label{Cav_opto}
\end{figure}

The dressed state picture discussed in the main text can also be used to describe the dynamics of cavity opto-mechanical systems.
It allows to recover in a simpler way the results obtained for these systems, such as the instabilities and the mechanical steady non-classical states.
The Hamiltonian for a driven opto-mechanical cavity in the rotating wave approximation reads \cite{Law_interaction_1994}
\begin{equation}
H_s = -\delta a^\dagger a + \omega_m b^\dagger b + g_O a^\dagger a (b^\dagger + b) + \Omega (a^\dagger + a) \ .
\end{equation}
Here $a$ ($b$) is the annihilation operator for the cavity (the oscillator), $g_O$ is the single-photon opto-mechanical coupling, $\Omega$ is Rabi frequency of the source and $\delta$ is the detuning of that source.
We follow the same method as described in the main text.
We first apply a LF transformation, defined as $U_{2} = e^{-g_O a^\dagger a (b^\dagger - b)/\omega_m}$, to diagonalize exactly the opto-mechanical coupling.
This leads to
\begin{equation}
U_2^\dagger H_s U_2 = -\delta a^\dagger a + \omega_m b^\dagger b - K (a^\dagger a)^2 + \Omega \left(a^\dagger e^{g_O(b^\dagger - b)/\omega_m} + a e^{-g_O(b^\dagger - b)/\omega_m} \right) \ ,
\end{equation}
where we introduced the Kerr term $K = g_O^2/\omega_m$.
For $\Omega = 0$, the eigenstates of the systems are $\ket{N,n}$, where $\ket{N}$ ($\ket{n}$) indicates the Fock state  of the cavity (the oscillator).
For $\Omega \neq 0$, we describe the laser interaction using a degenerate perturbation theory in $\Omega$.
For that, we consider a very weak laser intensity such that the cavity is populated at maximum by $1$ photon, and we then select the frequency of the laser for which $\delta = \omega_m - K + \epsilon$. 
In this way, the states $\ket{0,n}$ and $\ket{1,n+1}$ are nearly degenerated.
Note that when $g_O/\omega_m \lesssim 1$ the anharmonicity generated by the interacting term leads to photon blockade, preventing the population of higher photonic states.
We define the eigenstates $\ket{\pm,n}$ that are linear combinations of $\ket{0,n}$ and $\ket{1,n+1}$.
The system is then described by the master equation
\begin{equation}
\frac{d}{dt} \rho_s = \kappa \mathcal{D} \left( a e^{-g_O(b^\dagger-a)/\omega_m} \right) \rho_s + \gamma (1+n_B) \mathcal{D} (b) \rho_s  + \gamma n_B \mathcal{D} (b^\dagger)\rho_s  \ ,
\end{equation}
where $\kappa$ ($\gamma$) is the damping rate of the cavity (the oscillator), and $\rho_s$ the density matrix of the system in the LF basis.
Note that here we use the usual dissipators, and do not re-derive the whole master equation as we did in Appendix B.
This equation reduces to a Pauli rate equation in the secular regime, where $\sqrt{\epsilon^2 + 4 (\Omega W_{n,n-1})^2} \ll \kappa, \gamma$.
In that limit we derive the steady state of the system.
Figure \ref{Cav_opto}(a) shows the probability distribution of phonons for the strong opto-mechanical coupling $g_O = 0.8 \omega_m$.
Multiple maxima can be observed, as in the case of a driven TLS coupled to a mechanical oscillator.
Similarly, the sharpness of the first maximum resemble a Fock state, hence the negativity in the Wigner distribution associated [Fig.~\ref{Cav_opto}(b)]. 

%


\bibliographystyle{apsrev4-1}
\bibliography{Supplemental_Material}